\documentclass[iop]{emulateapj}
\usepackage{natbib}

\newcommand{\rms}{\mathrm{rms}}

\shorttitle{On Dark Peaks and Missing Mass}
\shortauthors{Clowe et al.}

\begin{document}


\title{On Dark Peaks and Missing Mass: A Weak Lensing Mass Reconstruction of the Merging Cluster System Abell 520 \footnote{Based on observations made with the NASA/ESA Hubble Space Telescope, obtained at the Space Telescope Science Institute, which is operated by the Association of Universities for Research in Astronomy, Inc., under NASA contract NAS 5-26555. These observations are associated with program \# 12253.} \footnote{This paper includes data gathered with the 6.5 meter Magellan Telescopes located at Las Campanas Observatory, Chile.}}


\author{Douglas Clowe}
\affil{Department of Physics and Astronomy, Ohio University, 251B Clippinger Labs, Athens, OH 45701}
\email{clowe@ohio.edu}

\author{Maxim Markevitch}
\affil{NASA Goddard Space Flight Center, Code 662, 8800 Greenbelt Rd., Greenbelt, MD 20706}

\author{Maru\v{s}a Brada\v{c}}
\affil{Department of Physics, University of California, One  Shields Avenue, Davis, CA 95616}

\author{Anthony H. Gonzalez and Sun Mi Chung}
\affil{Department of Astronomy, University of Florida, 211 Bryant
Space Science Center, Gainesville, FL 32611}

\author{Richard Massey}
\affil{Department of Physics, Durham University, South Road, Durham DH1 3LE, England}

\and

\author{Dennis Zaritsky}
\affil{Steward Observatory, University of Arizona, 933 N Cherry Ave, 
Tucson, AZ 85721}



\begin{abstract}
Merging clusters of galaxies are unique in their power to directly probe and place limits on the self-interaction cross-section of dark matter. Detailed observations of several merging clusters have shown the intracluster gas to be displaced from the centroids of dark matter and galaxy density by ram pressure, while the latter components are spatially coincident, consistent with collisionless dark matter. This has been
used to place upper limits on the dark matter particle self-inteaction cross-section of order 1 cm$^2$\,g$^{-1}$. The cluster Abell 520 has been
seen as a possible exception. We revisit A520 presenting new HST ACS mosaic images and a Magellan image set. We perform a detailed weak lensing analysis and show that the weak lensing mass measurements and morphologies of the core galaxy-filled structures are mostly in good agreement with previous works. There is however one significant difference -- we do not detect the previously claimed ``dark core" that contains excess mass with no significant galaxy overdensity at the location of the X-ray plasma.  This peak has been suggested to be indicative of a large self-interaction cross-section for dark matter (at least $\sim 5\sigma$ larger than the upper limit of $0.7 \mathrm{cm}^2\, \mathrm{g}^{-1}$ determined by observations of the Bullet Cluster). We find no such indication and instead find that the mass distribution of A520, after subtraction of the X-ray plasma mass, is in good agreement with the luminosity distribution of the cluster galaxies.  We conclude that A520 shows no evidence to contradict the collisionless dark matter scenario.
\end{abstract}





\section{Introduction}

Merging clusters of galaxies provide a unique opportunity to study properties of dark matter.
During a cluster merger, the cluster galaxies are effectively collisionless particles, slowed only by tidal
interactions, while the ionized X-ray emitting plasma clouds are affected by ram pressure as they pass
through each other.  The ram pressure causes the plasma clouds to slow down, and shortly after each 
collision in the merger process the X-ray plasma clouds will be found between the major concentrations
of cluster galaxies \citep[e.g.][]{RO97.1}.  Any dark matter present would be located in the vicinity of the
cluster galaxies, provided the dark matter does not a large self-interaction cross-section.
Because the X-ray plasma in a cluster makes up $\sim 12\%$ of the mass of a rich cluster \citep{AL02.1,VI06.1} while
the stellar mass in the cluster galaxies is less then $1\%$ of the mass of the cluster \citep{KO03.1}, one has
the situation that shortly after each collision the bulk of the baryonic matter is spatially displaced
from the the bulk of the total mass, and thus from the largest gravitational potentials in the cluster.  By comparing the
positions of the X-ray plasma clouds to the centers of the gravitational potential, as revealed using gravitational
lensing measurements, one can place measurements on the properties of dark matter and test alternative
theories of gravity, which one could use to replace some or all of the dark matter in galaxies and clusters.

The first such measurement of a merging cluster was performed on 1E0657-56 (aka ``The Bullet Cluster"), where
the gravitational potential of the two merging components were found to be spatially coincident with the cluster galaxies
but  significanctly displaced from the X-ray plasma clouds \citep{CL04.1,CL06.1}.  These observations
were later refined using strong lensing to place tighter constraints on the location and shape of the gravitational 
potential near the cluster cores \citep{BR06.1,BR09.1}, and were used to place constraints on the self-interaction
cross-section of dark mater \citep{MA04.1,RA08.1} as well as on any hypothetical ``5th force" that affects only
dark matter \citep{SP07.1,FA07.1}.  The Bullet Cluster has also been used to test alternative theories of
gravity, with the common result being that modified gravity models can reduce the amount of dark matter
needed in the system, but that the majority of the mass must still be in a dark, relatively non-collisional form
\citep{AN07.1,BR07.1,FE08.1}.  Several other merging clusters have since been found that provide similar results to those from the
Bullet Cluster.  These clusters are MACS J0025.4-1222 \citep{BR08.1}, A1758 \citep{OK08.1,RA12.1},  A2744 \citep{ME11.1}, A2163 \citep{SO12.1}, A754, A1750,
A1914, A2034, and A2142 \citep{OK08.1}.

A weak lensing study \citep[][hereafter M07]{MA07.1} of Abell 520 \citep[hereafter A520,][]{ACO}, at $z=0.199$ \citep{ST99.1}, finds instead
a large weak lensing signal in the location of the primary X-ray plasma cloud (location 3 in Fig.~\ref{kappamap}), well away from any large concentrations
of cluster galaxies, and no significant weak lensing signal in the location of one of the cluster galaxy concentrations, labeled as structure 5 in Fig.~\ref{kappamap}.  
A mass reconstruction by \citet{OK08.1} using part of the same imaging set as M07 did find significant mass in structure 5, but also found sufficient mass in location 3 to be consistent
with the M07 results.
More recently, this ``dark peak" has been confirmed by \citet[herafter J12]{J12.1} using a single passband imaging mosaic from the Wide Field Planetary Camera 2 (WFPC2) on the Hubble Space Telescope (HST).  The confirmation is not wholly independent as the WFPC2 mosaic data was combined with a ground based image from the Canada France Hawaii Telescope (CFHT) that was used in M07.  The majority of the signal on the locations and masses of the detected structures in the core of A520 does, however, come from the new WFPC2 data in the combined data set.  The confirmation of the dark peak in J12 is claimed at a $\sim 10\sigma$ detection level, with similar masses for the core structures as was measured in M07 except for a detection of significant mass at the cluster galaxy concentration of location 5, similar to the results of \citet{OK08.1}.
M07 and J12 provide several scenarios for how such a dark peak could arise,  such as a filamentary structure extending from the cluster, ejection of bright galaxies from a core during the merger process \citep[e.g.][]{SA07.1}, or a large self-interaction cross-section for
dark matter, so large that their quoted value lies beyond the 5-$\sigma$ upper limit on the cross-section
derived from the Bullet Cluster \citep{RA08.1}.  These results have also been used by several authors to argue in favor of
an alternative gravity model \citep{MO09.1,BE10.1}.

This is not the first time, however, that an apparently high-significance mass over-density that is not near a galaxy over-density has been found in weak lensing 
mass reconstructions.  In the cluster A1942, a mass over-density was found by \citet{ER00.1} roughly $7\arcmin$ away from the cluster core
and not near any significant galaxy concentrations.  In a separate case, \citet{MI02.1} found in a blank field STIS image a set of 11 galaxies arranged in a pattern
reminiscent of those seen in cores of massive clusters with multiple strongly lensed background galaxies.  In both cases, the significance
of the detections, as measured by the likelihood that a randomly chosen set of galaxies within the survey area would have a similar correlation in
their orientations on the sky, were sufficiently large that one would not expect to find such systems by chance.
However, in both cases, deeper observations resulted in the measurement of fainter galaxies that do not have the same correlated orientation, and therefore in the new analyses the mass over-densities either greatly diminished in amplitude (in the case of A1942 \citep{VL06.1}) or completely vanished (in the case of the STIS dark lens \citep{ER03.1}).

We present fully independent weak lensing observations of the merging cluster system A520 from a combined imagining data set from the Magellan 6.5m telescope in Chile and the Advanced Camera for Surveys (ACS) on HST.
These ground-based images are of much longer exposure times than those used in any of the previous weak lensing studies on this cluster and the ACS mosaic is deeper than the WFPC2 mosaic due to the higher throughput of ACS, and has three observed passbands for color selection of galaxies as compared to the monochromatic WFPC2 mosaic.   We investigate
whether this deeper data set confirms the existence of a significant mass over-density at the location of the X-ray plasma cloud.  The observations
are presented in \S\ref{sec:obs} and the weak lensing analysis in \S\ref{sec:wl}.  Discussion of the results are presented in \S\ref{sec:dis}, and we
summarize our conclusions in \S\ref{sec:con}.  Throughout this paper we assume a cosmological model with $\Omega_\mathrm{m} = 0.27$, 
$\Omega_{\Lambda} = 0.73$, $H_0 = 70$ km/s/Mpc and conventional gravity unless stated otherwise.

\section{Observations}\label{sec:obs}

\subsection{HST ACS Images}\label{sec:hst}

We obtained HST imaging with ACS on Feb 25-26 and Apr 6-7 2011 (HST Cycle 18 proposal
12253, PI Clowe). The new ACS data consist of four pointings in F435W,
F606W, and F814W. The corresponding exposure times are $2300\mbox{s}$
(1 orbit) in F435W and F606W, and $4600\mbox{s}$ (2 orbits) in F814W
per pointing. Each orbit was split into 4 dither positions, with a
large enough offset to cover the chip gap. 

Since the primary goal of this program is weak lensing analysis we
took special care when reducing and combining the images. During its
$\sim10$ years above the protection of the Earth's atmosphere, ACS has
accumulated significant radiation damage that has degraded its CCD
detectors. After each exposure, as photoelectrons are transferred
through the silicon substrate to the readout electronics, a certain
fraction is temporarily retained by lattice defects created by the
radiation damage, and released after a short delay \citep{JA01.1}. 
This effect is known as `Charge Transfer Inefficiency' (CTI)
and spuriously elongates the shapes of (in particular) faint galaxies
in a way that mimics weak gravitational lensing.

{ We have extended the CTI measurements of \citet{MA10.2} with a new
analysis of hot pixels in extragalactic
archival HST imaging taken before and after the A520 data.
Interpolating to the two epochs during which A520 data were obtained,
this analysis suggests that ACS observations taken in Feb 2011 (Apr
2011), 3283 (3324) days after launch suffer from 1.36747 (1.37949)
traps per pixel, and that the residual effective trap density after
correction is lower by a factor $20$.
There was insufficient data taken near that time to directly measure
the spurious shear in images obtained at that time. However,
extrapolating from the trap densities measured in HST COSMOS imaging
\citep{MA10.1}, this corresponds to a spurious shear before
correction of $\sim14\%$ spurious shear for faint ($26<m_{\mathrm{F814W}}<27$) galaxies furthest from the
readout register. This drops
rapidly to $\sim4.5\%$ by $25<m_{\mathrm{F814W}}<26$ and further at brighter magnitudes,
and falls linearly to zero as one approaches the readout register.}

Two independent pipelines have been developed to correct the image
trailing. Both use the iterative scheme of \citet{BR03.1} to move
electrons back, pixel-by-pixel, to where they belong. The first
pipeline, by \citet{MA10.1} (and updated for post-SM4 operations
by \citet{MA10.2}) is based around a physical model of charge capture and
release \citep{SH52.1,HA52.1}; the second, by \citet{AN11.1}, is built empirically from the observed trail
profiles. { Both methods have a demonstrated level of correction that leaves sub-percent
spurious shear residuals everywhere on the image at all magnitudes.}
We separately apply each of these pipelines to the ACS
imaging as the first step in data reduction.  Furthermore all images
taken with the ACS/WFC after Servicing Mission 4 show a row-correlated
noise due to the CCD Electronics Box Replacement. We correct for it
using the pyraf task {\tt acs\_destripe} \citep{GR10.1}.\footnote{\tt
  http://www.stsci.edu/hst/acs/software/destripe/}.  { The resulting weak lensing shear
measurements for the two CTI correction pipelines were consistent within $1\%$ of each other, in agreement
with our estimate of the expected residual shear from the CTI correction.  At this level there is minimal
effect on the weak lensing mass measurements presented herein.}

To stack the corrected data we use the Multidrizzle
\citep{multidrizzle} routine to align and combine the images. To
register the images we determine the offsets among the individual
exposures by extracting high S/N objects in the individual, distortion
corrected exposures. We use SExtractor \citep{BE96.1} and the IRAF
routine geomap to identify the objects and calculate the residual
shifts and rotations of individual exposures, which were then fed back
into Multidrizzle. We use square as the final drizzling kernel and an
output pixel scale of $0.05\arcsec$.  { The resulting images have
$5\sigma$ limiting magnitudes for galaxies, based on where
the number counts depart from an exponential growth function, of
$m_{\mathrm{F435W}} = 27.0$, $m_{\mathrm{F606W}} = 26.8$, and
$m_{\mathrm{F814W}} = 26.5$.}

\subsection{Magellan Optical Images}

We oboserved A520 with the IMACS camera on the Magellan Baade telescope during January 16-19, 2004.  
The camera was in the f/4 setup, resulting in a $0\farcs111$ arcsecond/pixel plate scale and $15\farcm4$ field of view.
During this time there were two significant problems with IMACS:  The atmospheric distortion corrector had not yet been delivered
to the telescope, and a problem with the CCD amplifiers created horizontal streaking in images after a saturated pixel was read.
The lack of the ADC caused the flat part of the focal plane to be much smaller than it was supposed to be, which resulted in only the
central $\sim 6\arcmin$ being in focus, and the image getting further out of focus the further one moves away from the center of the camera.
As a result, while many of the images were obtained with $\sim 0\farcs6$ seeing, they had $1\farcs0$ effective seeing at the edges of the image,  $\sim8\arcmin$ from
the center, and $1\farcs4$ seeing with noticeable coma in the corners of the images.  The horizontal streaking occurred in 4 of the 8 CCD chips, but 
only after highly saturated pixels were read out.  The magnitude of the streaks seemed to be dependent on both the total charge in the saturated pixels
and, oddly, the vertical position of the saturated pixel on the CCD -- in two of the chips, the streaking was very strong at the top and bottom of the chip,
but almost entire gone in the middle, despite being clearly caused by either the horizontal read-out register or the on-chip amplifier.  Due in part to our
limited data set and the large changes in the streak amplitude with both total charge and chip position, we were unable to find a good method of subtracting
the horizontal streaking.  We therefore left it in the images, being sure to mask any streaks prior to the creation of flat fields and removed any galaxies that overlapped
a streak from our weak lensing galaxy catalog.

We observed A520 in three passbands, Bessel $B$, $V$, and $R$, with single image exposure times of 5 minutes, chosen as a compromise between
minimizing the number of saturated stars on the images and minimizing the time lost to CCD readout.  Between each image,
we moved the telescope by $15\arcsec$ to fill in chip gaps and sample around bad pixels.  Our final integration times were
120 minutes in $R$ and 40 minutes each in $B$ and $V$.  Seeing varied between $0\farcs5$ and $0\farcs7$ in the $R$ images, $0\farcs7 - 0\farcs9$ in the
$V$ images, and $\sim 1\farcs0$ in the $B$ images.  Conditions were largely photometric, with stellar fluxes varying by only a few percent from image to image.

We performed image reduction by following the prescription for mosaic CCD reduction given in \citet{CL01.1}, doing bias subtraction with master bias frames, corrections by fitting
the overscan strip on each chip, and create flat fields by averaging together the science images with sigma-clipping after removing all detected objects from 
the images.  We register the images using a two step process of converting each CCD to a detector plane coordinate grid using a linear shift in x and y and a rotation
in the x--y plane.  We then map the detector plane onto the sky using a 7th order two-dimensional polynomial by comparing stellar positions to those in the USNO-B catalog
\citep{MO03.1}.
All of the images use the same CCD to detector plane conversion parameters, but the coefficients of the detector plane to sky conversion polynomial freely vary for each image. 
We therefore have 21 free parameters from the CCD to detector plane conversion and $36 \times n_{\mathrm{images}}$ free parameters from the polynomial coefficients, and
roughly $200 \times n_{\mathrm{images}}$ stellar positions to constrain the fit.  The resulting stellar positions have an average root mean squared (rms) of $0\farcs004$ compared to the same stars
in other images from this dataset, and a rms position difference of $0\farcs25$ when compared to the USNO, which is fairly typical of the positional uncertainties within the
USNO-B catalog \citep{MO03.1}.  One possible source of failure in this method is if the CCD chips in the camera are not sufficiently well aligned vertically, because then the detector-plane
to sky coordinate conversion can change too rapidly across the chip gap for the relatively low-order polynomial.  We test for
this by comparing the rms positions of stars that appear on more than one chip to those residing exclusively on a single chip and by looking for changes in the shapes of the PSF across chip gaps.  In both tests, we find no significant deviation that would indicate a vertical misalignment of the CCDs to a degree that would affect either the
image registration process or the subsequent weak lensing analysis.

Using the polynomials from the registration process, we map the images onto a common coordinate grid, preserving the $0\farcs111$ arcsecond/pixel plate scale and
orientation of the original images by using a triangular method with linear interpolation that preserves surface brightness and has been shown to not induce systematic changes
in object shapes for fractional pixel shifts \citep{CL00.1}.  We produce the final images by co-adding the registered images using a sigma-clipping algorithm to
detect and remove cosmic rays, while not clipping the centers or wings of stars.  The final images have FWHM  in the central $6\arcmin$ of $0\farcs63$ in $R$, $0\farcs75$ 
in $V$, and $1\farcs05$ in $B$, with increasing FWHM with distance from the center of the image.  { The $5\sigma$ limiting magnitudes for galaxies in the image centers in the final
images, as measured from where the number counts depart from an exponential growth function, are $m_B = 26.2$, $m_V = 25.8$, and $m_R=25.7$.  The $5\sigma$ limiting magnitudes
for galaxies at the edges of the images decrease by $\sim 0.2$ magnitudes due to the larger PSF size.}

\subsection{X-ray Images}

We have created an approximate projected gas mass map using a 0.8--4
keV X-ray image extracted from the archival Chandra 520 ks dataset
(Markevitch et al.\ in preparation). The X-ray emissivity at photon
energies $E\ll T_e$\/ depends very weakly on gas temperature and its
variations across the cluster. A520 is a $T_e=7$ keV cluster, and
Chandra has a peak of sensitivity at $E\simeq 1$ keV, which makes
the X-ray surface brightness in our energy band a good representation
of the projected X-ray emission measure, $EM\propto n_e n_p$.

To convert the projected emission measure to the gas mass requires
knowledge of the three-dimensional cluster geometry.  Unlike the
Bullet cluster \citep{MA02.1,CL06.1} that
appears to have a simple geometry, A520 is irregular and we cannot
make any plausible assumptions about its gas distribution along the
line of sight. An approach often used in such situations to obtain a
first-approximation gas mass map is to take a square root of the X-ray
brightness. As in \citet{RA12.1}, we attempt a slightly
higher level of accuracy by taking advantage of the fact that clusters
are centrally peaked and approximately spherically symmetric on large
scales. To do this, we first fit a spherically-symmetric
$\beta$-model to the X-ray radial brightness profile and create a
projected gas mass that corresponds to that model. This
zero-approximation mass map is then multiplied by a factor
$(S_X/S_\beta)^{1/2}$, where $S_X$ is the cluster surface brightness
and $S_\beta$ is the $\beta$-model image. To normalize this gas mass,
we use a $M_{\rm gas}-T$\/ relation from \citet{VI09.1},
which was derived from the Chandra gas masses and X-ray
temperatures, and the overall cluster temperature of 7.1 keV (Govoni
et al.\ 2004). Though the $M_{\rm gas}-T$\/ relation is derived for
relaxed clusters, hydrodynamic simulations indicate that it should not
be very different for mergers \citep[e.g.][]{NA07.1}. For the A520
temperature and redshift, the relation gives $M_{\rm gas}=7.9\times
10^{13}$ M$_\odot$ in a sphere of radius $r_{500}=1.10$ Mpc. We normalize
our map to have the same gas mass within the $r_{500}$ aperture as
that for the $\beta$-model.  { For a check, we have also tried a more direct (but also more noisy
for such irregular clusters as A520) estimate for the normalization
for our  $\beta$-model using the A520 Chandra spectrum from the
central $r=3\arcmin$ region and fitting it with the APEC spectral model \citep{APEC},
whose normalization gives the projected X-ray emission measure.  This
gives a gas mass $18\%$ higher than the above value within the same
sphere.}

Simulations, e.g., by Kravtsov et al. (2006) and Rasia et al. (2011),
indicate that even the extreme merging clusters, such as A520, follow
the $M-T$\/ relation with a scatter of about 20--25\% along the mass
axis, and the $M_{\rm gas}-M_{\rm tot}$\/ relation is even tighter.
Other errors in our analysis should be smaller, and we have assigned a
conservative 25\% error (68\% confidence) to the gas masses.  { The above $18\%$ difference
between our two gas mass estimates is well within this assumed
uncertainty.}

\section{Weak Lensing Analysis}\label{sec:wl}

\begin{figure*}
\plotone{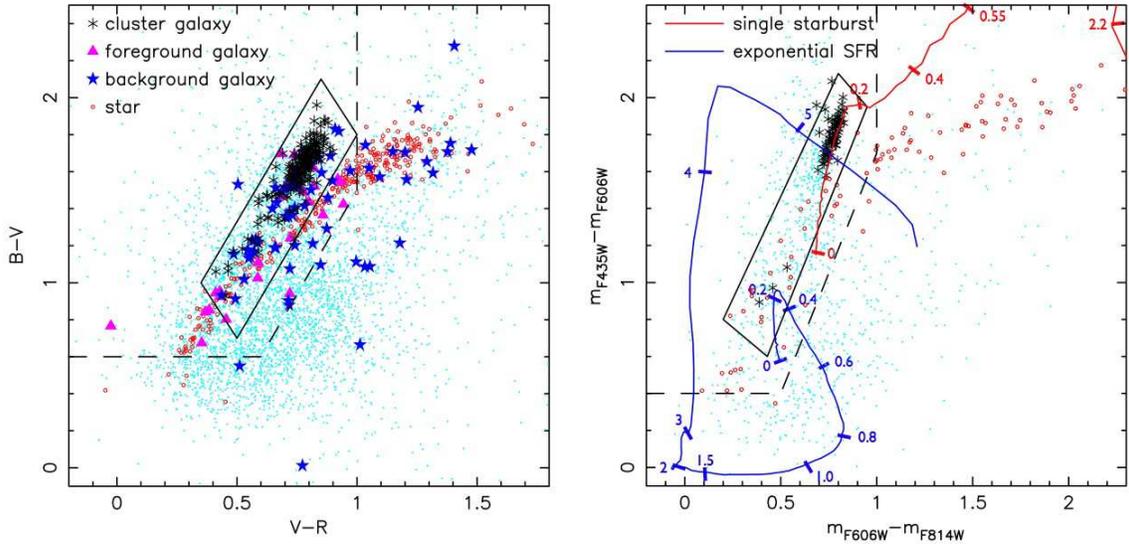}
\caption{\label{colsel} Shown above in the left (right) panel is the distribution in color-color space of galaxies in the Magellan (HST) images.  The cyan points show all of the galaxies with magnitude $R<24$ ($F606W < 24$), while the black, magenta, and blue asterisks denote the colors of spectroscopically confirmed cluster, foreground, and background galaxies respectively.  The red circles show colors of stars, rejected from the galaxy catalogs using size and central surface brightness criteria.  The solid box shows the selection for likely cluster galaxies used to perform the cluster luminosity measurements, while the dashed lines show the cuts used to remove likely cluster and foreground galaxies from the shear catalog (galaxies to the right and below the dashed line are kept).  The red and blue lines plotted on the right panel show theoretical colors for galaxies formed at $z=6$ with Salpeter initial mass function \citep{SA55.1} and present day solar metallicity with a single star-burst population (red) and an 10 Gyr exponential decay star formation rate (blue).  The tick marks and associated labels indicate where galaxies of a given redshift reside along these lines.  The color evolution models were generated using the EzGal software \citep{MA12.1} using the CB07 models.}
\end{figure*}

\subsection{Shear Measurement}\label{sec:shm}

We perform weak lensing analysis on the images with the goal of obtaining a two-dimensional distribution of the surface density in the cluster.  This is done by measuring second moments of the surface brightness to calculate an ellipticity for each galaxy, correcting this shape for smearing by the point spread function (PSF) to measure a shear, rejecting stars by size and central surface brightness, and rejecting likely cluster and foreground galaxies by color.  The methodology we use for the PSF correction is that of a modified KSB technique \citep{KSB}, details for which can be found in \citet{CL06.2}.  { Galaxies selected for the weak lensing analysis had a photometric S/N $> 10$ in the $R$ or F814W passbands, $R>22$ or $F814W > 21.5$, and did not have any bright neighbors near enough to significantly influence the second moment measurements (less than 3 times the sum of the scale radii of the galaxy and the neighbor).}  Weak lensing analysis is performed separately on the co-added Magellan image and on each of the four ACS pointings, then we combine the resulting shear catalogs to produce a final catalog.  The weak lensing measurements were performed with a modified version of the IMCAT\footnote{\tt http://www.ifa.hawaii.edu/\~kaiser/imcat} software package.

For the Magellan image, two additional defects in the image require added modification of the shear measurements.  The first is the horizontal streaking coming off of saturated stars in 4 of the CCDs.  As these streaks appear to be a change in the bias and/or gain of the readout amplifier, we do not trust the shapes of any objects in these regions, and therefore simply remove all objects intersecting any of these horizontal streaks in the image.  Due to the non-local nature of gravitational shear, this removal of galaxies will not bias our results, except to slightly increase the noise in the mass reconstructions in the vicinity of the removed galaxies.  

The second defect is that the images go out of focus at the edges of the image, which causes a strong change in the shape and size of the PSF as a function of radial position from the center of the image.  For the $R$-band image, from which we measure the galaxy shapes, the PSF size (FWHM) increases from $\sim 0\farcs 63$ in the center to $\sim 0\farcs 85$ at the edges and $\sim 1\farcs4$ in the corners.  Further, coma can be seen in the PSF in the corners of the image.  As a result, we restrict our shear measurements to a $8\arcmin$ radius from the center of the image and, instead of a straight size cut to separate stars and galaxies, we use a 7th order polynomial fit to the stellar half-light radius as a function of image position and rejected any object with a size smaller than $0\farcs1$ larger than the fit value at that location.  We supplement the stellar rejection by also rejecting objects with unusually high central surface brightness for its magnitude.  We also measure the KSB PSF correction terms ($P_{\mathrm{sh}}, P_{\mathrm{sm}},$ and stellar ellipticity) using a broad range of weighting function sizes, fit these as 7th order polynomials for image position variations, and use the fitted values for the PSF correction of a given galaxy based on its position and size.  To obtain the final $P_\gamma$ correction factor in the KSB technique, we divide the image up into four regions based on PSF size  to fit $P_\gamma$ as a function of galaxy size and ellipticity and reduce the significant noise present in the $P_\gamma$ measurement for each individual galaxy.   Using simulations with PSFs taken from the Magellan image, we found this technique systematically underestimates the measured shears by $\sim 13\%$ for a $0\farcs6$ PSF increasing to $\sim 15\%$ for a $0\farcs8$ PSF for the smallest galaxies in the simulations, decreasing to $\sim 10\%$ for galaxies significantly larger than the PSF size.  We determine this correction factor from fits to the simulation results based on the size of the PSF in the galaxy's location and the size of the galaxy.  

For the HST ACS images, we rejected stars using a size cut ($<0\farcs 0.081$ for $50\%$ encircled light radius) as well as rejecting objects with unusually high central surface brightness for their magnitude.  We again measure the KSB PSF correction terms for a range of weighting function sizes, fitted these using a 5th order polynomial for image position variations in each pointing, and use the fitted values matched to the galaxy size for correcting the PSF smearing.  We use the ACS-like STEP3 (http://www.roe.ac.uk/$\sim$heymans/step/cosmic\_shear\_test.html) simulations to calibrate the PSF corrections, finding a systematic underestimate of $\sim 8\%$ for the shear measurements, which was corrected for in the ACS measurements.  We perform the shear measurements independently for each of the three ACS passbands.

Because the HST and Magellan images observe galaxy populations with different redshift distributions, due mainly to loss of shape information on intrinsically smaller galaxies with the ground based PSF, and the strength of the shear measurements depends both on the mass of the lens and the redshifts of the background galaxies (see \S\ref{sec:mr}), before the shear measurements can be averaged between the two datasets we need to adjust the catalogs to have the same mean lensing depth.  Using external photometric redshift catalogs (see \S\ref{sec:mm} for details), we determined that the HST dataset would have a mean lensing signal in a given region that is $\sim 1.05$ times that of the Magellan image.  We therefore scale the shears measured in the Magellan image by 1.05 before combining with the HST catalog to create the final weak lensing catalog.

We compute weights for each galaxy in each data set by computing the inverse of the rms shear for nearby neighbors in significance and size space, with each data set showing that large, bright galaxies have a rms intrinsic shape of $\rms_g = 0.24$ per shear component in the F814W ACS passband, $0.26$ for the F606W ACS passband and the Magellan $R$-band, and $0.27$ for the F435W ACS passband.  Fainter and smaller galaxies have larger rms shear values, indicating increasing measurement errors for the second moments from sky noise and  PSF correction factors.  We therefore separate the rms shear values into two components, an intrinsic shape value chosen to be $\rms_{\mathrm{in}} = 0.24$ and a measurement value computed as $\rms_m = \sqrt{\rms^2_g - \rms^2_{\mathrm{in}}}$, and set a lower limit on $\rms_m = 0.05$ based on the spread in the $\rms_g$ values for the brightest and largest galaxies.  From these we create two weighting functions, $w_g = 1/\rms_g$ for weighting shears in the weak lensing mass reconstructions, and $w_m = 1/\rms_m$ 
for weighting the co-addition of shears for galaxies with multiple shear values in different ACS passbands, overlapping ACS pointings, and those galaxies located in both the ACS and Magellan images.  We compute a final weight for each galaxy by adding the $w_m$ values for each shear measurement in quadrature, taking the inverse to get a final $\rms_m$, and taking the inverse of $\rms_m$ added in quadrature with $\rms_{\mathrm{in}}$.

The final step in creating the weak lensing shear catalogs is to remove likely foreground and cluster galaxies from the galaxy catalogs.  This we do by using the {\it Hyperz} photometric redshift code \citep{BO00.1} to produce magnitudes in each observed passband for a range of galaxy templates from starbursts to passive ellipticals for $0<z<0.25$, adjusting these for the galactic extinction of the A520 field, isolating the regions in color-color space for the ACS and Magellan passbands for these galaxies, and removing all galaxies from the shear catalog that have colors, within photometric errors, which lie within the low-redshift galaxy color-color regions.  { To account for noise in the photometric measurements of fainter galaxies possibly moving foreground and cluster galaxies across the selection boundaries, we excluded all galaxies whose colors were within $1\sigma$ of the boundaries (a more stringent cut at higher $\sigma$ resulted in the loss of too many faint galaxies and a severe decrease in the S/N of the shear measurement).  These color-color cuts, shown in Fig.~\ref{colsel},} remove $\sim 40\%$ of the Magellan and $\sim 30\%$ of the ACS galaxies that otherwise were considered bright enough, large enough, and isolated enough to provide good shear measurements.  The final lensing catalog has a number density of galaxies of 22 per square arcsecond for the Magellan images and 56 for the ACS images, which result in rms shear per square arcminute of $0.036$ and $0.058$ for the regions around the core of A520 with and without ACS imaging respectively.  The number density of galaxies decreases by $\sim 15\%$, and rms shear increases by $\sim 10\%$, in the Magellan image as one approaches the edges of the image due to the increased PSF size.

\subsection{Mass Reconstruction}\label{sec:mr}

\begin{figure*}
\plotone{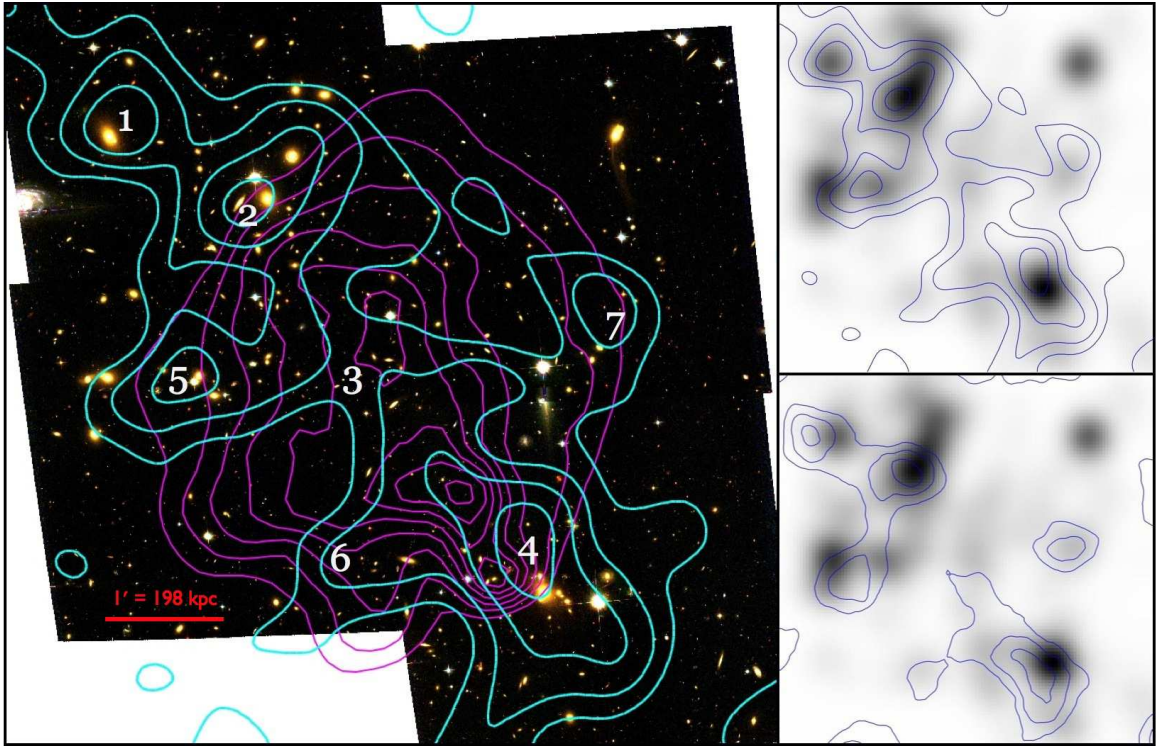}
\caption{\label{kappamap} Shown above in the left panel is a $7\arcmin \times 6\farcm75$ color composite from the HST ACS mosaic images with the weak lensing surface density reconstruction overlayed in cyan contours and the Chandra X-ray derived gas surface density in magenta contours.  The weak lensing contours show steps in surface density of $2\times 10^8$ M$_\odot/$kpc$^2$ ($\kappa$ steps of 0.056) above the mean surface mass density at the edge of the Magellan image ($\sim 1600$ kpc radius), and the gas mass contours show steps of $7.4\times 10^6$ M$_\odot/$kpc$^2$ with the outer contour starting at $4.4\times 10^7$ M$_\odot/$kpc$^2$.  The upper right panel shows the weak lensing contours superimposed on a smoothed cluster galaxy luminosity distribution in greyscale, with both the luminosity and surface density distributions smoothed by the same $\sigma = 60$ kpc Gaussian kernel.  The bottom right panel shows contours of the mass aperture statistic from the weak lensing data, with contours of steps of $1 \sigma$, superimposed on the cluster galaxy luminosity distribution.  Also labeled in the left panel are the regions of structures 1-6 identified in M07 and J12 as well as the new structure 7.}
\end{figure*}

The PSF corrected galaxy ellipticity measurements each provide independent measurements of the reduced shear $g$, where $g = \gamma/(1-\kappa)$ \citep[see][for formal derivations of the weak lensing concepts]{BA01.1}.  The shear $\gamma$ is the anisotropic change in the light distribution of the background galaxy caused by weak lensing, and the convergence $\kappa$ is the isotropic change in the background galaxy's light distribution.  The convergence is also the surface density of
the lens ($\Sigma$) scaled by a geometric factor ($\Sigma_{\mathrm{crit}}$) that depends on the angular diameter distances between the
observer, the lens, and the background galaxy being lensed.  Thus, to study the mass distribution of A520 with weak lensing, we need to convert our measured reduced shear data points to the convergence field of the lens, use an assumed cosmology and mean redshift for the background galaxies to estimate $\Sigma_{\mathrm{crit}}$, and scale the convergence field to a measurement of the surface density of the lens.

To convert the reduced shear measurements to the convergence field, we used the iterative technique of \citet{SE95.1}.  This technique is based on the KS93 algorithm \citep{KS93}, which uses that both $\gamma$ and $\kappa$ are second derivatives of the surface potential to combine derivatives of $\gamma$ to get derivatives of $\kappa$, which are then integrated to produce the convergence field to within an unknown integration constant, which is the mean value of $\kappa$ at the edge of the reconstructed field.  Because the input galaxy catalogs provide only a sparse sampling of the reduced shear field, the output $\kappa$ field needs to be smoothed to remove large noise spikes, in this case by convolution with a $\sigma=60 kpc$ Gaussian kernel.  The iterative technique is to initially assume $\kappa = 0$ across the reconstucted area, so $\gamma = g$, obtain a measurements of the $\kappa$ field, and then use this to perform a new correction of $\gamma = g \times (1 - \kappa)$.  After four iterations, we find the difference between the input $\kappa$ field and the output $\kappa$ field from the KS93 algorithm differ by less than $0.01\%$ of the input field, and stop the iteration.  For our combined catalog of 5903 background galaxies, the full set of iterations to produce a final two-dimensional (2-D) mass reconstruction takes only a few seconds.

The integration constant in each reconstruction is chosen by letting the mean value of $\kappa$ at the edge of the reconstructed area equal to the expected density of a cluster with the observed X-ray temperature.  For  our A520 data, this sets the convergence of the lens at a radius $r\sim 1500-1600$ kpc to $\kappa \sim 0.01$, { which is typical of clusters with X-ray temperatures of $\sim 7$ keV based on the $M_{500} - T_x$ relation of \citet{VI09.1}}.  While there is some dependence on the mass measured in the core region of A520 on this outer value, varying the convergence at this outer radius by $\pm 0.01$ results in a change in our measured substructure masses of less than $5\%$ absolute mass and $2\%$ relative mass ratios of structures, and has no discernible impact on the shapes or centroids of the substructure mass peaks.

The resulting $\kappa$ distribution is shown in contours overlayed on a color image constructed from the ACS mosaic data in the left-hand side of Fig.~\ref{kappamap}.  In the upper right-hand panel, the $\kappa$ contours are overlayed on a greyscale map of the luminosity distribution of cluster galaxies, selected by using the same color-color cuts that were used to exclude likely cluster galaxies from the weak lensing galaxy catalog, smoothed by the same sized Gaussian kernal as is the mass reconstruction.    We detect four primary mass concentrations (1, 2, 4, and 5 in Fig.~\ref{kappamap}), and see some evidence for excess mass in region 6 although none of our later tests would argue for a significant detection of an additional cluster substructure in this position.  We find no evidence of the mass overdensity in the dark peak region 3, and instead find a surface density distribution in that region which is in very good agreement with the underlying cluster galaxy light distribution.

In the bottom-right panel of Fig.~\ref{kappamap}, we also show the results of using a mass aperture ($M_{ap}$) statistic \citep{S96.1} on the weak lensing catalog.  The $M_{ap}$ statistic measures the $\kappa$ distribution convolved with a compensated filter, and has an advantage over the 2-D mass reconstruction in that it produces easily measurable errors.  The disadvantage is that the $M_{ap}$ statistic uses a more limited radial extent of the reduced shear measurements, and therefore has a lower signal-to-noise (hereafter S/N) in its measurement than the 2-D mass reconstruction.  To avoid having the negative portion of the compensated filter overlapping nearby structures, and thereby significantly decreasing the $M_{ap}$ signal, we used a 200 kpc outer radius for the statistic, and measured the values for centers distributed on a $100\times100$ grid across the cluster core region shown in Fig.~\ref{kappamap}.  We detect at $>2\sigma$ significance structures 1, 2, 4, and 5, detect structure 6 at only $1\sigma$ due in part due to its proximity to structure 4, and again find no detection of excess mass in the vicinity of the proposed dark peak.  We do find an additional structure that we label as structure 7 in Fig.~\ref{kappamap}, however it is significant only in the $M_{ap}$ measurements and not in the full mass reconstruction, which would be consistent with it being a local noise peak caused by a small number of highly elliptical galaxies in the vicinity.  We do not otherwise consider structure 7 in this paper.

\subsection{Bootstrap Resampling}\label{sec:bs}

\begin{figure*}
\plotone{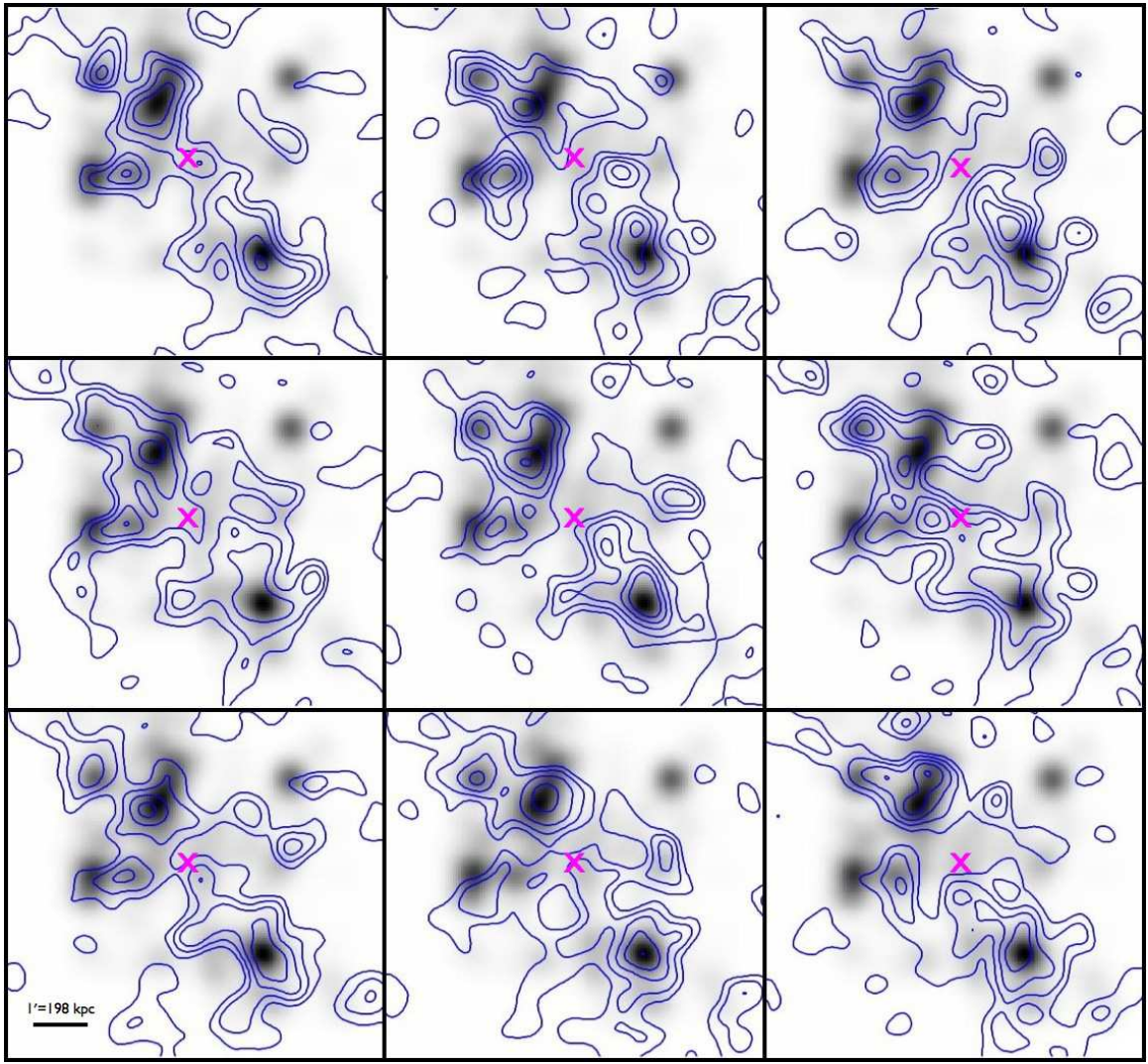}
\caption{\label{bootstrap} Shown above are nine randomly-selected weak lensing surface density reconstructions from the 100,000 bootstrap resampled catalogs used to measure errors in the weak lensing reconstructions superimposed on the cluster galaxy luminosity distribution.  The contour levels are identical to those in Fig.~\ref{kappamap}, and the magenta x shows the location of the dark peak of J12 (structure 3).  The middle right reconstruction shows a structure that is morphologically similar to and has similar mass as the dark peak in J12; such structures are found in $\sim 2\%$ of the bootstrap resampled reconstructions, and are the only reconstructions in which the column mass in the dark peak location agrees with that of J12.  Other reconstructions show smaller peaks near the dark peak, but have much less mass than reported by J12.}
\end{figure*}

Determining the errors on the 2-D mass reconstruction is more problematic than the errors on $M_{ap}$, as variations in the number density and magnitude of the intrinsic ellipticity of background galaxies cause the errors in both the enclosed mass, the mass centroid, and the mass structure shape to all vary by large amounts across the reconstructed area.  A common, but incorrect, method used to estimate these errors is to measure the rms shear and the mean density of the background galaxy catalog, and propagate these errors through the mass reconstruction algorithm obtaining an average noise level for the reconstruction.  The problem with this approach is that the $\kappa$ measurement for a given peak location is measured from the shear of galaxies in the catalog with an effective weighting of $\gamma/r$.  As $\gamma$ is largest and $r$ is smallest for galaxies immediately around the peak location, most of the weight in the $\kappa$ determination comes from a relatively small number of the nearest galaxies.  If any of these galaxies have an intrinsic ellipticity near the edges of the distribution function, the noise in that peak will be significantly larger than average.  

\begin{figure}
\plotone{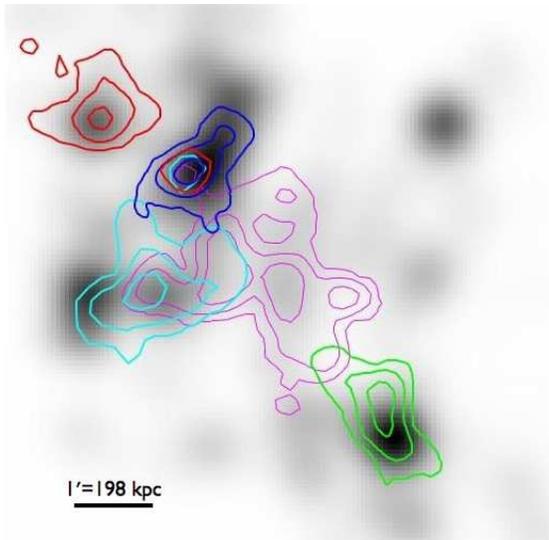}
\caption{\label{bscon} Shown above are contours enclosing the mass centroid position uncertainties for structures 1 (red), 2 (blue), 3(magenta), 4 (green), and 5 (cyan) superimposed on the cluster galaxy luminosity in greyscale.  The contours enclose the locations of the detected mass centroids in 100,000 bootstrap resamplings of the reduced shear catalog, and encompass 68\%, 95\%, and 99.7\% of the centroid measurements.}
\end{figure}

Another method to generate random noise fields that is commonly used is to preserve the position and total magnitude of the reduced shear measurement for each galaxy, but to apply a random orientation to each galaxy before performing the mass reconstruction.  After doing this enough times, one can then compute a rms of the noise field in each pixel of the reconstruction.  The problem with doing this in a field containing a massive cluster is two-fold: the cluster shear is still part of the measured reduced shear, so one would be significantly overestimating the combined intrinsic shape  and measurement noise for galaxies near the cluster core, and the average $\kappa$ in the random reconstructions will be 0, and thus the reconstruction will be misinterpreting the reduced shear to be shear, which also artificially enhances the level of the noise in the vicinity of the cluster core.  Thus, this method will give a lower limit on the significance of the detection of a structure in the weak lensing mass reconstruction, but in simulations of massive clusters we often find the lower limit can be as low as half of the true significance.  One can try to correct for this by using the smoothed mass reconstruction from the data to change the reduced shear into shear and then subtract off a shear field created from the mass reconstruction.  By doing this, however, you artificially reduce the level of the noise in the vicinity of noise peaks in the mass reconstruction, and thus overestimate the significance, in simulations often as much as a factor of 2, of the noise peaks.  One can therefore use this method to measure a minimum and maximum significance for the structures in the reconstruction, but the range between these two estimates is often quite large.

To properly measure how noise in the weak lensing catalogs affects the 2-D reconstruction, one needs to use a method that preserves the underlying reduced shear field while simulating the noise, which comes primarily from the intrinsic ellipticity of the background galaxies.
The method that we use to do this calculation is bootstrap resampling of the background galaxy catalog, in which one creates a new catalog with the same number of entries as in the original catalog, with each entry being a randomly selected member of the original catalogs and objects  are allowed to be selected more than once.  For a suitably large catalog, this results in the chance of any given object having an integer weight $m\ge 0$ to be $e^{-1}/m!$, and the chances of any group of $n$ galaxies not being in the new catalog to be $e^{-n}$.  Once the new catalog has been generated, a 2-D mass reconstruction can be measured from it, and by repeating the resampling as often as necessary, suitable statistics on the enclosed mass, mass centroid, and structure shapes can be measured.

We show nine randomly selected bootstrap resampled mass reconstructions in Fig.~\ref{bootstrap} as contours superimposed on the cluster luminosity
distribution in greyscale.  As would be expected, in general, the larger the mass peak in the original reconstruction, the smaller the changes in the relaltive size, position, and shape in the boostrap resampling reconstructions.  We see that most of the movements in the peak locations are well correlated with the shapes of the underlying galaxy distributions, consistent with a model where the galaxies are tracers of the dark matter mass distribution, and the observed peak locations are simply the largest noise peak in the vicinity of the structure core.  In most of the resamplings, we do not see a dark peak in the vicinity of location 3, although in about $2\%$ of the cases we do see a structure which one would identify as a mass peak not associated with cluster galaxies between locations 2 and 4.  One such example can be seen in the middle-right reconstruction of Fig.~\ref{bootstrap}.

Because the mass reconstructions have been smoothed to eliminate noise spikes from the reduced shear field, using a peak finder to detect the locations of the highest values of $\kappa$ within the various structures is equivalent to finding the mass centroid of the structure with a weighting function equal to the smoothing function used in the reconstruction.  In Fig.~\ref{bscon} we show the locations of the primary  mass
peaks for the various structures in 100,000 bootstrap resamplings, as contours that enclose 68\%, 95\%, and 99.7\%\ of the peak locations, which were detected as being the nearest significant peak to the structure in the original reconstruction.  Structure 2 is detected in all of the resamplings, and has a major axis in its centroid uncertainty distribution that agrees well with the NW-SE major axis of the distribution of cluster galaxies.  Structure 4 is also found in all of the resamplings, with a major axis in its centroid uncertainty running mostly N-S, and has a good agreement with the location of the cluster galaxies in about 1/3rd of the reconstructions.  This result suggests that the northern offset of structure 4 from the galaxy peak seen in the original reconstruction is likely just due to noise and not a significant feature of the merger. 

 In about $2\%$ of the reconstructions, structure 1 is not detected and instead the nearest significant peak is that of structure 2, otherwise the location of peak 1 is very centralized around the single giant elliptical galaxy in the region.  Structure 5 shows a major axis in its positional uncertainty that runs nearly E-W, in good agreement with the distribution of the cluster galaxies in this region.  The centroid of structure 5 overlays the brighter elliptical galaxies on the eastern end of the structure only $\sim 10\%$ of the time, while being detected in the vicinity of the dark peak location 3 about $\sim 2\%$ of the time, and not detected at all about $\sim 1\%$ of the time (when the nearest peak is that of structure 2).  In the $\sim 2\%$ of the cases where a peak is found in the vicinity of location 3, it is almost always ($\sim 90\%$) the case in which peak 5 is located at the extreme western end of its positional distribution rather than finding separate peaks in locations 3 and 5 as was seen in J12.  
 
 We also show in Fig.~\ref{bscon} the nearest peak to the dark peak location 3 from J12.  As opposed to the other 4 primary structures in the cluster, this distribution has multiple peaks in the centroid location distribution and is consistent with our finding in each reconstruction the nearest noise peak superimposed on a bridge structure traced out by the cluster galaxies.  When a peak with a mass equivalent to that in J12 is found near region 3 ($\sim 1\%$ of the reconstructions), $85\%$ of the time it is the case that structure 5 is at the extreme western end of its distribution and is the primary contributor to this increase in mass in region 3.  In the bootstrap resampling reconstructions, we do not see any indication of a dark peak in region 3 that is being suppressed in our original reconstruction by the chance projection of a handful of highly elliptical objects.
 
 To determine how these results are influenced by the size of the smoothing/weighting function used, we repeated them using  30 kpc and 120 kpc radius smoothing functions.  As expected for a white noise field superimposed on an underlying signal, we detect more mass peaks with the 30 kpc smoothing radius than for the 60 kpc smoothing, with the location of the most significant peak having a larger variation in position than with the 60 kpc smoothing radius.  For the 120 kpc smoothing radius, we get a slightly smaller ($\sim 80-90\%$ of the contour sizes seen in Fig.~\ref{bscon}) spread in the centroid locations when the structure is detected.  Structures 1 and 5 are not found $\sim 20\%$ of the time, instead appearing as extensions of structure 2, and structure 2 is not detected $\sim 5\%$ of the time.  Thus, the contours in Fig.~\ref{bscon} are likely to be slight overestimates of the true uncertainty in the mass centroid locations, but regardless of the smoothing functions we never detect a significant peak in the mass reconstruction at the location of structure 3 in more than $5\%$ of the bootstrap reconstructions.
 
 \subsection{Mass Measurements}\label{sec:mm}

Instead of asking whether A520 has a dark peak, a more direct question is whether there is excess mass in the vicinity of the reported location of the dark peak.  To do this, we measure a column mass enclosed within a given radius by simply integrating over the $\kappa$ values within the given radius around a chosen center in the 2-D mass reconstructions, and do the same in the bootstrap resampled reconstructions to look at the distribution of errors in the measurement.  For easy comparison, we use the same aperture size used in J12, 150 kpc, which results in non-overlapping mass integration regions around the peaks except for a 9 kpc overlap between peaks 3 and 5 and a 50 kpc overlap between peaks 4 and 6.  The mass of the X-ray plasma and integrated cluster luminosity in each region are computed using the cutout regions from the X-ray mass and cluster light images described earlier.  

To convert the integrated $\kappa$ values to column masses, we need to assume a value for $\Sigma_{\mathrm{crit}}$.  Because we do not have data in enough passbands to measure reliable photometric redshifts for the background galaxies, we use photometric redshifts from other imaging data sets for the magnitude range of the background galaxies.  From these, we calculate the inverse of the mean value of $\Sigma^{-1}_{\mathrm{crit}}$, using weights based on the F814W magnitude to mimic the weights used in the weak lensing measurements, that such a redshift distribution would have if it were to be lensed by a $z=0.2$ cluster.   The two photometric redshift datasets we used were the COSMOS field catalog of \citet{IL09.1}, for which the photometry is mainly ground-based Subaru data and thus a good  match to the Magellan image, and the UDF catalog of \citet{Coe06}, which has entirely space based photometry, and is thus a good match to the HST mosaic.  From these catalogs, we determined mean values of $\Sigma_{\mathrm{crit}}$ of $3.4 \times 10^{9}$ M$_\odot/$kpc$^2$ for the HST mosaic and $3.6 \times 10^{9}$ M$_\odot/$kpc$^2$ for the Magellan image.  As we scaled the Magellan shears to compensate for this prior to coaddition of the catalogs, we adopt the UDF value for converting $\kappa$ to surface density.  The mass ratios of the various apertures are insensitive to the adopted value of $\Sigma_{\mathrm{crit}}$, and show only minor variations when we change the scaling factor between the Magellan and HST catalogs.

\begin{figure*}
\plotone{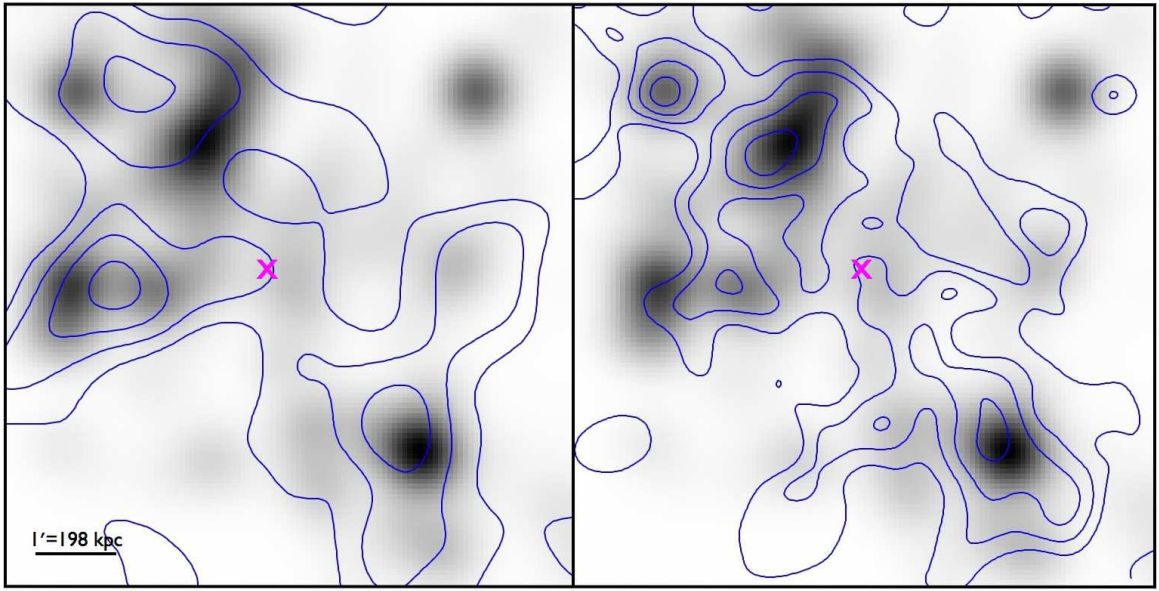}
\caption{\label{indivkappa} Shown above are the weak lensing surface density reconstructions using only the galaxies detected the Magellan image (left) and HST mosaic (right) as contours superimposed on the cluster galaxy luminosity in greyscale.  The Magellan reconstruction has been smoothed by a 
82 kpc Gaussian kernel, while the HST reconstruction was smoothed by a 26 kpc Gaussian kernel.  The magenta x shows the location of the dark peak in J12.}
\end{figure*}

The resulting mass and luminosity measurements for the six structures detected in our mass reconstruction and the dark peak location can be found in Table 1.  We also provide the errors for the lensing masses, determined from the variation in masses measured in the bootstrap resampled reconstructions.  We do not provide errors for the gas mass or cluster luminosity measurements, as the random errors between the measurements for the different locations are dwarfed by the error in the weak lensing masses.  We also list in Table 1 the mass-to-light ratio, computed after subtraction of the X-ray gas mass, and its error for each of the peaks.  There are likely systematic errors in all three measurements, however the systematic errors in the cluster luminosity measurements (e.g.~galactic dust, intracluster light) will only change the absolute scale of the mass-to-light measurements and not the relative values of the peaks.  Some non-cluster galaxies are likely to have been included by the color selection process, however in all 7 regions the majority of the galaxy light is emitted by red-sequence galaxies.  Using the same color selections on the COSMOS and UDF catalogs discussed above suggests that interlopers should be contributing $\sim 3 \times 10^9$ L$_\odot$ (calculated by converting flux to luminosity by placing all galaxies are at the cluster redshift) on average in each aperture, or less than $2\%$ of the measured luminosity.  Therefore fluctuations of a factor of a few in the interloper population among the apertures will produce minimal contamination of mass-to-light ratios.  The systematic errors in the weak lensing (e.g.~chosen value of $\Sigma_{\mathrm{crit}}$, PSF correction) are relatively small ($<10\%$) and while the systematic errors in the gas mass measurement could be quite large ($\sim 25\%$) the gas mass is sufficiently small compared to the weak lensing mass in all of the peaks that combined they will cause the relative mass-to-light ratios among the peaks to vary by less than the random errors.  

\begin{deluxetable*}{cllcccc}
\tabletypesize{\scriptsize}
\tablecaption{Mass Reconstruction Substructure Properties ($r < 150$ kpc)}
\tablewidth{0pt}
\tablecolumns{7}
\tablehead{\colhead{Substructure} & \colhead{RA} & \colhead{Dec} & \colhead{Column Mass} & \colhead{Luminosity} & \colhead{Gas Mass} & \colhead{M/L} \\ 
\colhead{} & \colhead{(${h:m:s}$)} & \colhead{(${\degr:\arcmin:\arcsec}$)} & \colhead{($h_{70}^{-1} 10^{13} M_\sun$)} & \colhead{($h_{70}^{-2} 10^{11} L_{z\sun}$)} & \colhead{($h_{70}^{-5/2} 10^{13} M_\sun$)} & \colhead{($M_\sun/L_{z\sun}$)} } 

\startdata
P1 & 04:54:19.60 & +02:57:49.09 & 3.03$\pm$0.69 & 2.43 & 0.25 & 114$\pm$28 \\
P2 & 04:54:14.84 & +02:57:06.25 & 4.08$\pm$0.73 & 4.16 & 0.40 & 88$\pm$18 \\
P3 & 04:54:11.25 & +02:55:37.28 & 2.26$\pm$0.75 & 1.38 & 0.69 & 114$\pm$54 \\
P4 & 04:54:04.57 & +02:53:58.60 & 4.64$\pm$0.63 & 3.11 & 0.50 & 133$\pm$20 \\
P5 & 04:54:17.11 & +02:55:30.09 & 3.00$\pm$0.77 & 2.66 & 0.44 & 96$\pm$29 \\
P6 & 04:54:09.61 & +02:53:55.90 & 3.03$\pm$0.66 & 1.15 & 0.65 & 207$\pm$57 \\
\enddata
\tablecomments{M/L is calculated after subtraction of the gas mass.}
\end{deluxetable*}

From the detected mass structure (1, 2, 4, and 5), we obtain a weighted mean mass-to-light ratio of $108\pm 24$ M$_\odot/$L$_\odot$, which is in excellent agreement with the mass-to-light ratio measured for the region of the dark peak ($114\pm54$ M$_\odot/$L$_\odot$).  For structures $1-5$ (we leave out 6 due to the large overlap with the structure 4 aperture), the hypothesis that all of the structures share the same mass-to-light ratio gives a reduced $\chi^2 = 0.75$.  If structure 4 is assumed to be a noise peak superimposed on the underlying mass distribution and instead we measure the mass-to-light centered on the cluster galaxy luminosity peak, the weak lensing mass decreases by only $\sim 2\%$ while the cluster luminosity increases by $\sim 17\%$, decreasing its mass-to-light ratio from $133\pm20$ to $111\pm17$, and the reduced $\chi^2$ of the constant mass-to-light hypothesis to $0.43$.  None of the other structures are misaligned with the underlying light distribution enough to significantly change the reduced $\chi^2$ by centering them on their cluster luminosity centroids.  The mass-to-light ratio of structure 6 is much higher than the others, but is still within $2\sigma$ of the mean.

{ Another method of measuring mass within a given radius, and that used by J12, is aperture densitometry \citep{FA94.1,CL00.1}.  The two traditional statistics measure the mean $\kappa$ within radius $r_1$ minus the mean $\kappa$ within a given annular region, with the difference being how the subtractive annular region is defined:}
\begin{eqnarray*}
\zeta(r_1) = \bar{\kappa}(r\leq r_1) - \bar{\kappa}(r_1<r\leq r_{\mathrm{max}}) \\
= {2 \over 1-r_1^2/r_{\mathrm{max}}^2} \int_{r_1}^{r_{max}} \langle \gamma_T \rangle d\ln r
\end{eqnarray*}
\begin{eqnarray*}
\zeta_c(r_1) = \bar{\kappa}(r\leq r_1) - \bar{\kappa}(r_2<r\leq r_{\mathrm{max}}) = 2 \int_{r_1}^{r_2} \langle \gamma_T\rangle d\ln r \\ + {2 \over 1-r_2^2/r_{\mathrm{max}}^2}
\int_{r_2}^{r_{\mathrm{max}}} \langle \gamma_T\rangle d\ln r,
\end{eqnarray*}
{ with $r_{\mathrm{max}}$ being the maximum radius used in the measurement, $r_2$ is the inner radius of the subtractive aperture for $\zeta_c$, $\gamma_T$ is the portion of the shear measurement oriented tangential to the chosen center, and the angular brackets indicate azimuthal averaging.
In practice, one can bin the tangential shear measurements in bins of constant logarithmic radius change ($d\ln r$) and convert the integrals in the above equations to summations over the bins.  This allows for an easy calculation of the errors, with $\sigma^2_\zeta = (\sum_{\mathrm{bin}} 4 (d\ln r)^2 \rms^2_g/n_{\mathrm{bin}})/(1 - r_1^2/r_{\mathrm{max}}^2)$, where $n_{\mathrm{bin}}$ is the number of galaxies in a given bin.  
One problem with this statistic, however, is that it is assuming one has measured the shear $\gamma$ instead of the reduced shear $g$, and thus overestimates the mean density when $\kappa$ is not small compared with 1.  This can be corrected using an iterative technique, similar to that used above in the 2-D mass reconstructions, of converting the current estimate of $\bar{\kappa} (r)$ from $\zeta_c$ to $\kappa (r)$, assuming the annular region is at large enough radius for the subtractive element to be small, using this estimate of $\kappa(r)$ to convert $g$ to $\gamma$ and recalculate $\zeta_c$.  As with the 2-D reconstruction, within 4-5 iterations the difference between successive iterations of $\zeta_c$ are small compared to the uncertainty in the measurement and thus the iteration can be stopped.  This will increase the error in the statistic due to the added uncertainty in the value of $\kappa$ used to corrected the reduced shear, however we have performed simulations of iterative aperture densitometry measurements around clusters with masses similar to A520 and find this additional error is small, usually increasing the measurement error by less than $5\%$ compared to that calculated above. }

\begin{deluxetable*}{ccccc}
\tabletypesize{\scriptsize}
\tablecaption{Aperture Densitometry Substructure Properties ($r < 150$ kpc)}
\tablewidth{0pt}
\tablecolumns{5}
\tablehead{\colhead{Substructure} & \colhead{$\zeta$ Column Mass} & \colhead{M/L ($\zeta$)} & \colhead{$\zeta_c$ Column Mass} & \colhead{M/L ($\zeta_c$)} \\ 
\colhead{} & \colhead{($h_{70}^{-1} 10^{13} M_\sun$)} & \colhead{($M_\sun/L_{z\sun}$)} & \colhead{($h_{70}^{-1} 10^{13} M_\sun$)} & \colhead{($M_\sun/L_{z\sun}$)} }
\startdata
P1 & 2.33$\pm$0.77 & 95$\pm$35 & 2.81$\pm$0.671 & 99$\pm$27 \\
P2 & 3.45$\pm$0.73 & 62$\pm$15 & 4.16$\pm$0.67 &  70$\pm$13 \\
P3 & 2.01$\pm$0.73 & 125$\pm$68 & 2.84$\pm$0.64 &  150$\pm$44 \\
P4 & 4.71$\pm$0.76 & 113$\pm$20 & 5.59$\pm$0.68 & 123$\pm$17 \\
P5 & 2.48$\pm$0.70 & 89$\pm$31 & 3.17$\pm$0.66 &  102$\pm$25 \\
P6 & 2.95$\pm$0.78 & 242$\pm$82 & 3.68$\pm$0.68 &  224$\pm$50 \\
\enddata
\tablecomments{M/L is calculated after subtraction of the gas mass.}
\end{deluxetable*}

{  We list in Table 2 the mass measurements within a 150 kpc radius using aperture densitometry for both statistics for each structure, in both cases using $r_{\mathrm{max}} =1500$ kpc  and  $r_2 = 1150$ kpc for $\zeta_c$.  We also show the resulting mass-to-light ratios; the measurements of the cluster luminosity in these regions are slightly different than those given in Table 1 as we use unsmoothed luminosity measurements corrected for subtraction the luminosity density of the same annular regions used in the aperture densitometry measurements.  Overall the results compare well, within errors, with the measurements from the 2-D mass reconstruction given the expected increase in the masses in $\zeta_c$ due to the lack of smoothing of the aperture densitometry measurements, and the reduction in $\zeta$ due to the subtraction of the mean $\kappa$ of the reconstruction region.  The exception is that of structure 3, for which we would expect the lack of smoothing to have little effect on the mass measurement, but instead is $\sim 25\%$ higher than the mass reconstruction value.
These results give a reduced $\chi^2$ for the first 5 peaks of 1.2 for $\zeta$ and 2.16 for $\zeta_c$, which reject the constant mass-to-light ratio hypothesis at $70\%$ and $93\%$ confidence levels respectively.  In both cases the main driver for the higher $\chi^2$ compared to that from the 2-D mass reconstruction is the lower mass-to-light ratio of structure 2 rather than the higher mass-to-light ratio of structure 3.  For all three mass measurement techniques, however, a constant mass-to-light ratio hypothesis for all of the structures in the cluster cannot be excluded at a level larger than $2\sigma$.}

For completeness, we also measure the total mass of the cluster using the iterative aperture densitometry technique at a radius of 700 kpc from the centroid of the cluster galaxy luminosity distribution, which fully contains the structures seen in Fig.~\ref{kappamap} but is small enough to still have a reasonably large weak lensing S/N measurement, to measure a column mass of $5.1\pm 0.7 \times 10^{14}$ M$_\odot$, in good agreement with the values given by M07 and J12.   Assuming that the cluster outside the core can still be modeled with a NFW profile \citep{NFW}, using a fixed concentration $c = 3.5$ gives a measurement of $M_{200} = 9.1\pm1.9 \times 10^{14}$ M$_\odot$.  Varying the concentration between 2 and 5 results in a variation in $M_{200}$ that is about half the error level of the weak lensing measurements.  The NFW profiles have $M_{500} = 6.1\pm1.3 \times10^{14}$ M$_\odot$, in good agreement with the X-ray derived $M_{500} = 6.7\pm1.0 \times 10^{14}$ M$_\odot$, calculated by using $T_x = 7.1\pm 0.7$ keV \citep{GO04.1} and the $M_{500} - T_x$ relation of \citet{VI09.1}.

\subsection{Strong Lensing}\label{sec:sl}

\begin{figure}
\plotone{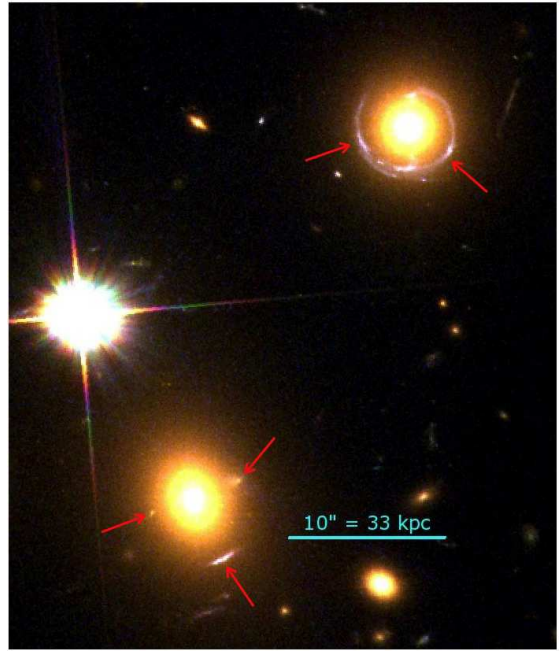}
\caption{\label{stronglens} Shown above in a color composite from the HST ACS images are the only two cases of strong lensing we detect in Abell 520.  Both sets of lenses are around giant elliptical galaxies located in structure 2, with the strong lensing features marked by red arrows.  Redshifts of the arcs are currently unknown.}
\end{figure}

{ One signal that could indisputably confirm the presence of the dark peak would be a strongly lensed galaxy in the vicinity of region 3.  A careful search for objects with large length-to-width ratio (giant arcs) and for objects with similar morphologies and colors (multiply imaged arclets) around region 3 in the ACS mosaic does not, however, reveal any obvious strongly lensed galaxies.  In fact, in the entire cluster system, the only two cases of strong lensing  that we find are both being caused primarily by individual cluster galaxies rather than by a cluster core.  These lenses are the two brightest elliptical galaxies in structure 2, and are shown in Fig.~\ref{stronglens}.  The likelihood of having such galaxy scaled lenses is known to be enhanced by the presence of a nearby cluster core \citep[see][]{Kn11.1}, however degeneracies between the shape and strength of the galaxy's gravitational potential and that of the cluster prevent us from using these strong lenses to place additional constraints on the presence or absence of a mass peak in region 3.   Given the cluster mass distribution's likely strong lensing cross-sections and the observed distribution, sizes, and redshifts of background galaxies in the field, it is not surprising that we do not detect any significant signs of strong lensing by the cluster cores.}

\section{Discussion}\label{sec:dis}

\begin{figure}
\plotone{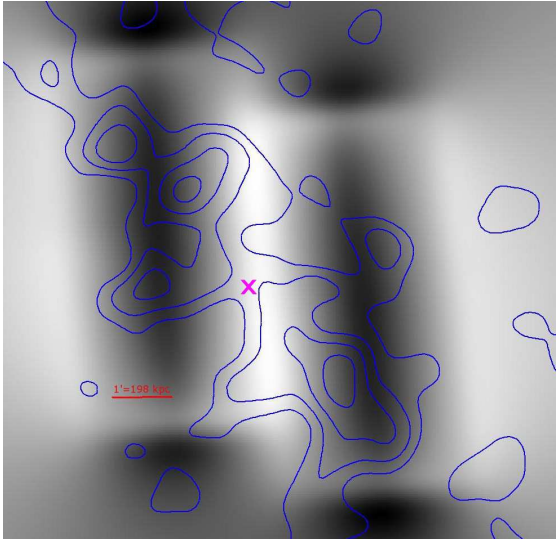}
\caption{\label{CTI} Shown above in greyscale is what the $\kappa$ distribution of the mass reconstruction area would look like if the only signal was spurious ellipticity caused by the CTI of the ACS camera.  The mass reconstruction from Fig.~\ref{kappamap} is shown as blue contours, with the magenta x indicating the location of the dark peak in J12.  The greyscale is set such that the white regions are relatively unaffected by the CTI, while the dark regions are the most affected.  With the
applied CTI correction, the extreme values in the CTI only mass reconstruction are  $\kappa \sim \pm 0.008$, which is $\sim 1/7$th the level of the first contour shown in the mass reconstruction overlay.}
\end{figure}

Comparison of our results described above to those in J12 reveals that  the weak lensing mass measurements and morphologies of the core structures are mostly in good agreement.   There are, however, three major differences between the studies: (1) the amount of light we measure from cluster galaxies in location 3 is twice that measured in J12, (2) despite using better data, our uncertainties of the masses in the weak lensing measurements are $1.5-1.7$ times those of J12, and (3) our weak lensing column mass in location 3 is $\sim 60-70\%$ of that meaured by J12, depending on which measurement technique is used.  Below we discuss possible reasons for these differences and potential ways to reconcile the two results.

There are a number of differences between the integrated galaxy luminosities in our Table 1 to those in J12.  The two largest are that ours are measured for the observed F814W passband (roughly a restframe $R$ passband), while the J12 luminosities were measured for a rest-frame $B$ passband, and our luminosities were measured from a luminosity image that had been smoothed by the same amount as the mass reconstructions, while the J12 luminosities are measured as aperture cutouts without any smoothing.  The smoothing decreases the measured luminosities in all of the structures reported in Table 1, but those of structures 2 and 4 are decreased by a larger fraction, $30\%$ and $35\%$ respectively, than those of structures 1, 3, 5, and 6, which range from $3-8\%$.  The column masses of the structures, however, show a similar increase when measured both on unsmoothed mass reconstructions and using the aperture densitometry technique.   As a result, while the measured mass-to-light ratio for structures 2 and 4 decrease relative to structure 3 in the unsmoothed measurements, it is only a $\sim 20\%$ effect, which is not large enough to cause structure 3's mass-to-light ratio to be significantly different from the other structures.  Changing from the F814W luminosities to F435W luminosities makes the luminosity ratio of structure 3 to structures 2 larger and therefore moves the mass-to-light ratio of structure 3 closer to that of structure 2, because all of the bright galaxies in structure 2 are red-sequence cluster ellipticals, while in structure 3 there are a number of luminous blue spiral galaxies.  Looking at the selection of galaxies in our color selected catalogs reveals that all of the bright galaxies in structures 2 and 4  are included in the catalog while several bright spiral galaxies were excluded by the color selection process in structure 3.  The inclusion of these galaxies would significantly increase the luminosity of structure 3 compared to structures 2 and 4.  If we further reduce the size of the color selection region to include only cluster red-sequence galaxies, we do reduce the luminosity of structure 3 compared to structure 2, but only by $\sim 10\%$.  As such, the largest luminosity ratio between structure 2 and 3 we can measure is 4.1:1, compared to the 3.0:1 in Table 1, and can not reproduce the 5.3:1 and 6.9:1 ratios of J12 and M07 respectively.  Thus, the majority of the difference between the mass-to-light ratios in the dark peak location between our results and those in J12 and M07 is not due to the weak lensing mass measurement, but comes from the difference in the luminosity measurements for galaxies at the location of structure 3.

For the error levels in the weak lensing mass measurements in Tables 1 and 2, the two primary determining factors in the data are the rms shear measurement level and the area used to make the shear measurements.  For the measurements in J12, the rms shear per square arcminute { using their weighting scheme} is 0.034 for the HST/WFPC2 images and 0.056 for the ground based CFHT image (Jee, priv comm), which are very comparable to our 0.036 from the HST/ACS images and 0.056 from the Magellan image { using our weighting scheme} (see \S\ref{sec:shm}).  The ground-based image used in J12 is larger than that used here, however the effective weight for each galaxy in the mass measurement is $\gamma/r \propto r^{-2}$, so the galaxies missing from our smaller field have low weight in the mass measurements for the 150 kpc apertures used.  Simulations using the measured rms shear values indicate that doubling the size of the ground based image while preserving the rms shear level would decrease the size of the error in the mass measurements in Table 1 by only $\sim 10\%$ of their current values.  

In addition, J12 use a much larger number density of background galaxies in their measurements from the WFPC2 images than we do from the ACS images (92 galaxies per sq.~arcminute versus our 56), and do not apply any color selection to their { galaxies fainter than $R=24$, but instead remove those galaxies with known redshifts that are part of the cluster and those in the red sequence which are brighter than $R=24$}.  As a result, while they have likely excluded all of the cluster giant ellipticals, many cluster and foreground dwarf galaxies are likely still in their background galaxy catalogs { as the majority of the galaxies in their catalogs would be at $R>24$}.  With the color selection, we reject galaxies with an average number density of 24 per sq.~arcminute (although higher towards the cluster cores than at the image edges), and using only these galaxies in a weak lensing mass reconstruction shows mostly noise with only a small trace of the cluster.  This suggests that the majority of these excluded galaxies are likely cluster members or foreground galaxies, and the inclusion of these would decrease the S/N of the weak lensing signal.  The remaining number density of 12 galaxies per sq.~arcminute difference between the two catalogs are likely galaxies fainter than our minimum detection significance cut, therefore having low weights in the weak lensing shear measurements and contributing little to the mass reconstruction.  { By comparison, the \citet{MI02.1} STIS dark peak had a number density of 16 galaxies per sq.~arcminute for the bright sample with the strong spurious alignment signal, and a number density of 77 galaxies per sq.~arcmin. giving the weaker alignment signal.}  Overall, the inclusion of faint cluster and foreground galaxies will increase the value of $\Sigma_{\mathrm{crit}}$ needed to convert $\kappa$ measurements into a surface mass, which is consistent with the J12 adopted $\Sigma_{\mathrm{crit}} = 4.1\times 10^9$ M$_\odot/$kpc$^2$ versus our $\Sigma_{\mathrm{crit}} = 3.6\times 10^9$ M$_\odot/$kpc$^2$.  This higher value of $\Sigma_{\mathrm{crit}}$ gives a higher error in the surface mass for a given error in $\kappa$.  Simulating the reported error levels for the shear measurements in J12 with the reported $\Sigma_{\mathrm{crit}}$ suggests that they should have error levels on the order of $\sim 7\times 10^{12}$ M$_\odot$ rather than the $\sim 4\times 10^{12}$ M$_\odot$ level that they report, and thus their significance levels are less than $60\%$ of what they state.

The most surprising difference between our results and those of J12, however, is the similarity of the weak lensing measurements for all parts of the cluster except that in the vicinity of region 3.  When an error is made in the analysis of a weak lensing data set, the resulting mass reconstruction normally differs from the true mass reconstruction across the entire field, and not a localized difference in one small portion of the mass reconstruction, especially if the difference in that one area is large compared to the noise level in the reconstructions.
Both J12 and M07 overstated the significance of their detections of this structure because they compared their measurements to a 0 mass level in that region instead of to a model of a constant mass-to-light ratio across the cluster.   We argue that our mass measurement of $2.3 \times 10^{13}$ M$_\odot$ for this region is a better baseline for comparison as it has a similar mass-to-light ratio as the other structures, and this would give a significance of the mass overdensity in this region of  $1.9\sigma$ for the M07 measurement and $4\sigma$ for the J12 measurement, using their quoted errors, or around $2.3\sigma$ using our estimated error level for their measurement discussed above.  The significance of the difference in the mass measurements for this region between our result and the J12 result is larger than the $2.3\sigma$ level, however, as one of the primary sources of noise, the intrinsic ellipticity of the background galaxies used in the weak lensing shear measurement, is largely in common between the two measurements.  Any deviation in shear measurements between the two catalogs would therefore either have to be an error in one of the measurement techniques, or a much more significant inherent alignment in the shapes of the galaxies used in J12 that are not included in our background galaxy catalogs.  At this level of difference, there are plenty of potential sources of error in the weak lensing measurements to explain how two such different reconstructions can be drawn from the same underlying data.

One potential source of error in a weak lensing mass reconstruction is that arising from an incorrect PSF smearing correction.  This is cited in M07 to explain the difference in their reconstruction of the Subaru image compared with an analysis in a early version of \citet{OK08.1}, which Okabe and Umetsu agreed with in the final version of their paper.  { It is unlikely an error in PSF correction is a factor for this case.  Both groups have tested their methodology on simulations and recover shears to better than a $1\%$ accuracy with known PSFs, well below the level that would create or remove such a large structure in the mass reconstruction.   Also, both groups have multiple image sets from different telescopes on which they have measured similar looking structures, making the likelihood of the presence or absence of a given structure being caused by a poorly measured PSF in that region small.}  Both J12 and M07 find a mass structure in region 3 using two ground based image sets from different telescopes and, for J12, a HST WFPC2 mosaic.  We show in Fig.~\ref{indivkappa} separate mass reconstructions for the Magellan ground based image and the HST ACS mosaic, neither of which exhibit a structure in region 3.  Because the different imaging sets have very different intrinsic PSF ellipticity distributions, it would be highly unlikely to make mistakes in each shear measurement such that they combine to give the same false weak lensing mass structure.

Another potential source of error is the treatment of the charge transfer inefficiency (CTI) in the HST images.  Both WFPC2 and ACS suffer from CTI, which is caused primarily by electron traps being created in the silicon detectors by high energy cosmic rays.  As these traps effectively shuffle charge to trails behind a bright object, they impart a fake, correlated ellipticity in the stars and galaxies that can mimic a weak lensing signal.    The WFPC2 CTI shape was measured by J12 and while they do find a potential residual in the location of the dark peak location, the residual is more than an order of magnitude below what would be needed to cause the observed structure.  For the ACS images, we are confident that CTI is not playing a significant role in the galaxy and PSF shape measurement process for two reasons.  The first is that we created two different sets of images using different CTI correction mechanisms (see \S\ref{sec:hst}) and measure nearly identical mass reconstructions.  { The second is that the induced ellipticity pattern from the ACS CTI when used in the 2-D mass reconstruction technique would create a four quadrant positive $\kappa$ signal as well as significant $\kappa$ peaks both north and south of the HST imaging area which is not seen in our A520 mass reconstruction (see Fig.~\ref{CTI}).  If both of the CTI correction techniques were ineffective in removing the induced shear signal, correction for this would not significantly change the mass measurement for region 3 as our tiling strategy for ACS placed it in a region with no CTI.  However, it would reduce the masses of all of the other structures by $25-40\%$, placing these masses in disagreement with that measured by J12 and that which we measure from only the Magellan data.   This would also shift the centroids of the weak lensing peaks for structures 1, 2, 4, and 5 away from the black bars in Fig.~\ref{CTI}, which would cause significant offsets between the mass and galaxy locations in these structures.  The reduction in the CTI effect by a factor of 20 from the correction techniques (\S~\ref{sec:hst}) will reduce the CTI caused $\kappa$ features by also roughly a factor of 20, which would have a maximum change of $\sim 1/7$th the value of the first contour of the overlays and change the measured masses in Table 1 by order $1-2\%$ for all but structure 3, which would still be unaffected by the CTI.}

A third potential source of error is a sudden change in the mean shear at the boundary of the ACS images due to differences in the redshift distributions of the background galaxies used in the ACS images compared to those in the Magellan image.  This would cause a sudden change in the mean $\Sigma_{\mathrm{crit}}$ that, as the mass reconstruction algorithm assumes $\Sigma_{\mathrm{crit}}$ is constant across the field, would be interpreted as a feature caused by the surface mass distribution of the cluster.  In simulations, such a mismatch usually causes a plateau of surface density with a size and shape similar to the ACS mosaic region, but can occasionally due to the coupling of random noise in the shear measurements and this systematic change at the boundary alter the relative masses and locations of the structures in the cluster core seen in the mass reconstructions.  We already attempted to correct for this effect by scaling the shear measurements of the ground based image to match the depth of the ACS image prior to combining the catalogs.  To test that this is not causing a problem, we have recombined the space-based shear catalog with both an unscaled ground-based catalog and one with twice the correction we originally used, and with neither of the two resulting catalogs does the mass reconstruction have any structure at the dark peak location (3).  We note that this type of rescaling of the ground-based shear measurements by a constant fraction is not strictly correct, as we are really measuring the reduced shear and both $\kappa$ and $\gamma$ will change with a variation in $\Sigma_{\mathrm{crit}}$.  As a final test, we used the original mass reconstruction to separate the $\gamma$ and $\kappa$ components of the reduced shear in the ground based image catalog, scaled each separately, and recombined them to get a scaled reduced shear catalog.  Combining this catalog with the ACS catalog also made no discernible change in the masses or centroids of the structures observed in the resulting mass reconstruction.

A more likely cause of the difference between the two reconstructions is not an error in the weak lensing measurements at all, it is simply a difference in which galaxies are selected to have their shapes measured and included in the shear field.  Because the intrinsic galaxy shapes are the largest random error for reasonably bright galaxies, different sets of galaxies will change the noise field in the reconstructions.  It is this source of noise that is reproduced in the bootstrap resampling methodology of \S\ref{sec:bs}.  As can be seen in Fig.~\ref{bootstrap}, we do see a structure near location 3 in about $2\%$ these bootstrap resampled reconstructions, which is consistent with our excluding the mass measured by J12 in this region at a $2.3\sigma$ level.   The additional cluster and foreground galaxies in the J12 catalogs discussed above could be a source for the additional shear needed to produce a peak in location 3, either through intrinsic or purely random alignments of the intrinsic shapes of the galaxies such as was seen in the ``dark lens" of the STIS fields in \citet{MI02.1}.  To know for certain the cause, however, will require a direct galaxy by galaxy comparison of the two catalogs to determine if the difference is within the shear measurements for galaxies in common between the two catalogs or from the extra galaxies used in the J12 catalog.

There is one feature in common in the mass reconstructions which is seemingly at odds with a CDM model of the merger, however.  East of structure 5 are several bright elliptical galaxies which have spectroscopic redshifts \citep{YE96.1, CA96.1, PR00.1} that identify them as cluster members, but neither our nor the J12 mass reconstruction show any appreciable mass overdensity around these galaxies despite them being several times more luminous that those of structure 5.  The bootstrap resampled catalogs do have peak 5 moving far enough eastward to be associated with these galaxies, but only $\sim 10\%$ of the time.  If real, this lack of mass could be an example of ejection of bright galaxies from cores during a merger event.  However, as these galaxies are located at the edge of both the ACS and WFPC2 mosaics, it is possible that a boundary effect problem, as discussed above, could be causing a drop in surface density at this location.  In the Magellan only reconstruction seen in Fig.~\ref{indivkappa}, we do find structure 5 to have an extended mass tail across this location, although the centroid is still more consistent with the fainter ellipticals of location 5.

\section{Conclusions}\label{sec:con}

Using a new multi-color HST ACS mosaic and previously unpublished Magellan image set, we performed a  weak lensing mass reconstruction on the merging cluster A520.  The mass structures in the reconstruction show excellent agreement with the distribution of light from cluster galaxies after subtraction of the mass of the intra-cluster X-ray plasma.  While the masses we measure for the cluster overall and all of the cluster substructures containing galaxies are in good agreement with previous weak lensing measurements in J12, we do not detect the mass overdensity spatially coincident with the X-ray plasma cloud that was found in both M07 and J12.  We measure a total mass in this region consistent with a constant mass-to-light ratio across the cluster, and exclude the additional mass in the central region at a $\sim 98\%$ confidence level.  { Using an aperture densitometry measurement instead of the mass reconstruction results in a slightly higher mass for the dark peak region, it still excludes the mass from J12 at a $93\%$ confidence level.   The mass measurements from the 2-D mass reconstruction are consistent with a constant mass-to-light model, while the mass aperture show marginal evidence for a departure from constant mass-to-light ratio, primarily caused by a lower mass-to-light ratio measurement than average in structure 2.}  We also find that the significances for the dark peak structure were overstated in both M07 and J12 as they calculated the significance by comparing their measured mass to a mass of 0 in the center of the cluster, and their significances of detection are $<2\sigma$ and $\sim 2.3\sigma$ respectively when measured compared to our constant mass-to-light ratio model.  

We have considered several potential causes for the discrepancy in the central region between the various mass reconstructions while still having good agreement in the rest of the cluster core.  We suggest that the most likely explanation is an inherent alignment in the galaxies that were included in the J12 shear measurements but excluded from ours.  Regarding the discrepent mass-to-light ratio, we find that both M07 and J12 have significantly lower cluster luminosity measurements in the region of the ``dark peak" than our measurements, and this difference in cluster luminosity is responsible for the majority of the difference in the mass-to-light ratios for the central region between the studies.

We identify one structure on the eastern edge of the HST image which has bright elliptical galaxies that are known to be part of the cluster for which neither we nor J12 obtain a significant amount of mass.  However, we do detect mass in this region in a mass reconstruction shape measurements from the Magellan image.  We are uncertain if the lack of mass in this region is an aspect of the merger or an edge effect from the HST mosaic.  

The overall mass structure that we measure for A520 is in good agreement with a constant mass-to-light ratio, and therefore with collisionless cold dark matter --- similar to the conclusions drawn from all other well-studied merging clusters.   Deriving a quantitative upper limit on the dark matter self-interaction cross-section from A520 will require additional kinematic information
and detailed modeling of this merging system.

\acknowledgments

We wish to thank  Alexie Leauthaud and James Taylor for useful discussions on the shear measurement errors seen in COSMOS, and Dan Coe for providing a copy of the UDF photometric redshift catalog matched to our observed passbands.  We also wish to thank James Jee, Andisheh Mahdavi, and Henk Hoekstra for both discussions on possible differences between the observed shear catalogs and additional information on their techniques that helped rule out several possibilities for differences between the mass reconstructions.
Support for program \# 12253 was provided by NASA through a grant from the Space Telescope Science Institute, which is operated by the Association of Universities for Research in Astronomy, Inc., under NASA contract NAS 5-26555.



{\it Facilities:} \facility{Magellan}, \facility{HST}.

\clearpage

\clearpage

\clearpage

\clearpage







\begin{thebibliography}{}
\bibitem[Abell, Corwin, \& Olowin(1989)]{ACO} Abell, G.~O., Corwin, H.~G., \& Olowin, R.~P. 1989, \apjs, 70, 1
\bibitem[{Allen} {et~al.}(2002)]{AL02.1} {Allen}, S.~W., {Schmidt}, R.~W., \& {Fabian}, A.~C. 2002, \mnras, 334, L11
\bibitem[Anderson \& Bedin(2011)]{AN11.1} Anderson, J. \& Bedin, L.~R. 2011, \pasp, 122, 1035
\bibitem[{Angus} {et~al.}(2007)]{AN07.1} Angus, G.~W., Shan, H.~Y., Zhao, H.~S., \& Famaey, B. 2007, \apjl, 654, L13
\bibitem[Bartelmann \& Schneider(2001)]{BA01.1} Bartelmann, M. \& Schneider, P. 2001, \physrep, 340, 291
\bibitem[Bekenstein(2010)]{BE10.1} Bekenstein, J.~D.  2010, in Particle Dark Matter: Observations, Models and Searches, ed. G. Bertone (Cambridge U. Press, Cambridge), 95
\bibitem[Bertin \& Arnouts(1996)]{BE96.1} Bertin, E. \& Arnouts, S. 1996, \aaps, 317, 393
\bibitem[Bolzonella, Miralles, \& Pell{\'o}(2000)]{BO00.1} Bolzonella, M., Miralles, J.-M., \& Pell{\'o}, R. 2000, \aap, 363, 476
\bibitem[{Brada\v c} {et~al.}(2006)]{BR06.1} {Brada{\v c}}, M., {Clowe}, D., {Gonzalez}, A.~H., {Marshall}, P., {Forman}, W., {Jones}, C., {Markevitch}, M., {Randall}, S., {Schrabback}, T., \& {Zaritsky}, D. 2006, \apj, 652, 937
\bibitem[{Brada\v c} {et~al.}(2008)]{BR08.1} {Brada{\v c}}, M., {Allen}, S.~W., {Treu}, T., {Ebeling}, H., {Massey}, R., {Morris}, R.~G., {von der Linden}, A., \& {Applegate}, D. 2008, \apj, 687, 959
\bibitem[Brada\v c {et~al.}(2009)]{BR09.1} {Brada{\v c}}, M., {Treu}, T., {Applegate}, D., {Gonzalez}, A.~H., {Clowe}, D., {Forman}, W., {Jones}, C.,  {Marshall}, P., {Schneider}, P., \& {Zaritsky}, D. 2009, \apj, 706, 1201
\bibitem[Bristow(2003)]{BR03.1} Bristow, P. 2003, Instrument Science Report STIS-2003-001 (astro-ph/0310714)
\bibitem[Brownstein \& Moffat(2007)]{BR07.1} Brownstein, J.~R. \& Moffat, J.~W. 2007, \mnras, 382, 29
\bibitem[Carlberg {et~al.}(1996)]{CA96.1} Carlberg, R.~G., ,Yee, H.~K.~C., Ellingson, E., Abraham, R., Gravel, P., Morris, S., \& Pritchet, C.~J. 1996, \apj, 462, 32
\bibitem[{Clowe} {et~al.}(2000)]{CL00.1} {Clowe}, D., {Luppino}, G.~A., {Kaiser}, N., \& {Gioia}, I.~M. 2000, \apj, 539, 540
\bibitem[{{Clowe} \& {Schneider}(2001)}]{CL01.1} {Clowe}, D. \& {Schneider}, P. 2001, \aap, 379, 384
\bibitem[{{Clowe}, {Gonzalez}, \& {Markevitch}}(2004)]{CL04.1} {Clowe}, D., {Gonzalez}, A., \& {Markevitch}, M. 2004, \apj, 604, 596
\bibitem[{Clowe} {et~al.}(2006)]{CL06.1} Clowe, D., {Brada{\v c}}, M., {Gonzalez}, A.~H., {Markevitch}, M., {Randall}, S.~W., {Jones}, C., \& {Zaritsky}, D. 2006, \apjl, 648, L109
\bibitem[{{Clowe} {et~al.}(2006)}]{CL06.2} {Clowe}, D. {et~al.} 2006, \aap, 451, 395
\bibitem[Coe {et~al.}(2006)]{Coe06} Coe, D., Benitez, N., Sanchez, S.~F., Jee, M., Bouwens, R., \& Ford, H. 2006, \apj, 132, 926
\bibitem[Erben {et~al.}(2000)]{ER00.1} {Erben}, T., {van Waerbeke}, L., {Mellier}, Y., {Schneider}, P., {Cuillandre}, J.-C., {Castander}, F.~J., \&{Dantel-Fort}, M. 2000, \aap, 355, 23
\bibitem[Erben {et~al.}(2003)]{ER03.1} Erben, T., Miralles, J.-M., Clowe, D., Schirmer, M., Schneider, P., Freudling, W., Pirzkal, N., Fosbury, R.~A. E., \& Jain, B. 2003, \aap, 410, 45
\bibitem[Fahlman {et~al.}(1994)]{FA94.1} Fahlman, G., Kaiser, N., Squires, G., \& Woods, D. 1994, \apj, 437, 56
\bibitem[{Farrar} \& {Rosen}(2007)]{FA07.1} Farrar, G.~R. \& Rosen, R.~A. 2007, \prl, 98, 171302
\bibitem[{Feix}, {Fedeli}, \& {Bartelmann}(2008)]{FE08.1} Feix, M., Fedeli, C., \& Bartelmann, M. 2008, \aap, 480, 313
\bibitem[Foster {et~al.}(2012)]{APEC} Foster, A.~R., Ji, L., Smith, R.~K., \& Brickhouse, N.~S. 2012, \apj, submitted, arXiv:1207.0576
\bibitem[Govoni {et~al.}(2004)]{GO04.1} Govoni, F., Markevitch, M., Vikhlinin, A., van Speybroeck, L, Feretti, L., \& Giovannini, G. 2004, \apj, 604, 695
\bibitem[Grogin {et~al.}(2010)]{GR10.1} Grogin, N.~A., Lim, P.~L., Maybhate, A., Hook, R.~N., \& Loose, M. 2010, in 2010 Space Telescope Science Institute Calibration Workshop, Baltimore, MD, 54
\bibitem[Hall(1952)]{HA52.1} Hall, R. 1952, Phys. Rev., 88, 139
\bibitem[Ilbert {et~al.}(2009)]{IL09.1} Ilbert, O., et al. 2009, \apj, 690, 1236
\bibitem[Janesick(2001)]{JA01.1} Janesick, J.~R. 2001, Scientific Charge-coupled Devices (Bellingham, WA: SPIE Press)
\bibitem[Jee {et~al.}(2012)]{J12.1} Jee, M.~J., Mahdavi, A., Hoekstra, H., Babul, A., Dalcanton, J.~J., Carroll, P., \& Capak, P. 2012, \apj, 747, 96
\bibitem[{{Kaiser} \& {Squires}(1993)}]{KS93} {Kaiser}, N. \& {Squires}, G. 1993, \apj, 404, 441
\bibitem[{Kaiser}, {Squires}, \& {Broadhurst}(1995)]{KSB} {Kaiser}, N., {Squires}, G., \& {Broadhurst}, T. 1995, \apj, 449, 460
\bibitem[Kneib \& Natarajan(2011)]{Kn11.1} Kneib, J.-P. \& Natarajan, P. 2011, \aapr, 19, 47
\bibitem[{Kochanek} {et~al.}(2003)]{KO03.1} {Kochanek}, C.~S., White, M., Huchra, J., Macri, L., Jarrett, T.~H., Schneider,  S.~E., \& Mader, J. 2003, \apj, 585, 161
\bibitem[Koekemoer {et~al.}(2002)]{multidrizzle} Koekemoer, A.~M., Fruchter, A.~S., Hook, R.~N., \& Hack, W. 2002, in The 2002 HST Calibration Workshop, Baltimore, MD, 339
\bibitem[Mancone \& Gonzalez(2012)]{MA12.1} Mancone, C.~L. \& Gonzalez, A.~H. 2012, \pasp, 124, 606
\bibitem[Mahdavi {et~al.}(2007)]{MA07.1} Mahdavi, A., Hoekstra, H., Babul, A., Balam, D.~D., \& Capak, P.~L. 2007, \apj, 668, 806
\bibitem[Markevitch {et~al.}(2002)]{MA02.1} Markevitch, M., Gonzalez, A.~H., David, L., Vikhlinin, A., Murray, S., Forman, W., Jones, C., \& Tucker, W. 2002, \apj, 567, L27
\bibitem[Markevitch {et~al.}(2004)]{MA04.1} {Markevitch}, M., {Gonzalez}, A.~H., {Clowe}, D., {Vikhlinin}, A., {Forman}, W., {Jones}, C., {Murray}, S., \& {Tucker}, W. 2004, \apj, 606, 819
\bibitem[Massey {et~al.}(2010)]{MA10.1} Massey, R., Stoughton, C., Leauthaud, A., Rhodes, J., Koekemoer, A., Ellis, R., \& Shaghoulian, E. 2010, \mnras, 401, 371
\bibitem[Massey(2010)]{MA10.2} Massey, R. 2010, \mnras, 409, L109
\bibitem[Merten {et~al.}(2011)]{ME11.1} Merten, J., et al. 2011, \mnras, 417, 333
\bibitem[Miralles {et~al.}(2002)]{MI02.1} Miralles, J.-M., Erben, T., H{\"a}mmerle, H., Schneider, P., Fosbury, R. A. E., Freudling, W., Pirzkal, N., Jain, B., \& White, S. D. M. 2002, \aap, 388, 68
\bibitem[Moffat \& Toth(2009)]{MO09.1} Moffat, J.~W. \& Toth, V.~T. 2009, \mnras, 397, 1885
\bibitem[Monet {et~al.}(2003)]{MO03.1} Monet, D.~G., et al. 2003, \apj, 125, 984
\bibitem[Nagai, Vikhlinin, \& Kravtsov(2007)]{NA07.1} Nagai, D., Vikhlinin, A., \& Kravtsov, A.~V. 2007, \apj, 655, 98
\bibitem[{Navarro}, {Frenk}, \& {White}(1995)]{NFW} {Navarro}, J.~F., {Frenk}, C.~S., \& {White}, S.~D.~M. 1996, \apj, 462, 563
\bibitem[Okabe \& Umetsu(2008)]{OK08.1} Okabe, N. \& Umetsu, K. 2008, \pasj, 60, 345
\bibitem[Proust {et~al.}(2000)]{PR00.1} Proust, D., Cuevas, H., Capelato, H.~V., Sodre, L, JR., Tome Lehodey, B., Le Fevre, O., \& Mazure, A. 2000, \aap, 355, 443
\bibitem[Ragozzine {et~al.}(2012)]{RA12.1} Ragozzine, B., Clowe, D., Markevitch, M., Gonzalez, A.~H., \& Brada\v c, M. 2012, \apj, 744, 94
\bibitem[{Randall} {et~al.}(2008)]{RA08.1} {Randall}, S.~W.,  {Markevitch}, M., {Clowe}, D., {Gonzalez}, A.~H., \& {Brada{\v c}}, M. 2008, \apj, 679, 1173
\bibitem[Roettiger, Loken, \& Burns(1997)]{RO97.1} Roettiger, K., Loken, C., \& Burns, J.~O. 1997, \apjs, 109, 307
\bibitem[Sales {et~al.}(2007)]{SA07.1} Sales, L.~V., Navarro, J.~F., Abadi, M.~G., \& Steinmetz, M. 2007, \mnras, 379, 1475
\bibitem[Salpeter(1955)]{SA55.1} Salpeter, E.~E. 1955, \apj, 121, 161
\bibitem[Schneider(1996)]{S96.1} Schneider, P. 1996, \mnras, 283, 837
\bibitem[Seitz \& Schneider(1995)]{SE95.1} Seitz, C. \& Schneider, P. 1995, \aap, 297, 287
\bibitem[Shockley \& Read(1952)]{SH52.1} Shockley, W. \& Read, W. 1952, Phys. Rev., 87, 835
\bibitem[Soucail(2012)]{SO12.1} Soucail, G. 2012, \aap, 540, A61
\bibitem[Springel \& Farrar(2007)]{SP07.1} Springel, V. \& Farrar, G.~R. 2007, \mnras, 380, 911
\bibitem[Struble \& Rood(1999)]{ST99.1} Struble, M.~F. \& Rood, H.~J. 1999, \apjs, 125, 35
\bibitem[{Vikhlinin} {et~al.}(2006)]{VI06.1} Vikhlinin, A., Kravtsov, A., Forman, W., Jones, C., Markevitch, M., Murray,  S.~S., \& {Van Speybroeck}, L. 2006, \apj, 640, 691
\bibitem[{Vikhlinin} {et~al.}(2009)]{VI09.1} Vikhlinin, A., et al. 2009, \apj, 692, 1033
\bibitem[{von der Linden} {et~al.}(2006)]{VL06.1} von der Linden, A., Erben, T., Schneider, P., \& Castander, F.~J. 2006, \aap, 454, 37
\bibitem[Yee, Ellingson, \& Carlberg(1996)]{YE96.1} Yee, H.~K.~C., Ellingson, E., \& Carlberg, R.~G. 1996, \apjs, 102, 269
\end{thebibliography}
\end{document}